\documentclass{aa}

\usepackage{graphicx,color}
\usepackage{grffile}
\usepackage[T1]{fontenc}
\usepackage[utf8]{inputenc}
\usepackage{lmodern}
\usepackage{txfonts}
\usepackage{caption}
\usepackage{float}

\usepackage{fourier}
\DeclareMathAlphabet{\mathcal}{OMS}{cmsy}{m}{n}
\SetMathAlphabet{\mathcal}{bold}{OMS}{cmsy}{b}{n}

\usepackage[colorlinks,bookmarks]{hyperref}
\definecolor{linkblue}{rgb}{0,0,0.8}
\definecolor{linkgreen}{rgb}{0,0.5,0}
\hypersetup{linkcolor=linkblue, citecolor=linkgreen, urlcolor=linkblue}

\definecolor{valecol}{rgb}{0,0.5, 1.}

\definecolor{davidcol}{rgb}{0.5,0, 0.5}

\definecolor{tiagocol}{rgb}{0.5, 0.5, 1.0}

\def\d{{\rm d}}

\newcommand{\aperp}{a_{\perp }}
\newcommand{\aperpo}{a_{\perp_0}}
\newcommand{\apar}{a_{\parallel}}
\newcommand{\Hperp}{H_{\perp}}
\newcommand{\Hperpo}{H_{\perp_0}}
\newcommand{\Hpar}{H_{\parallel}}

\newcommand{\Mhalo}{{\ifmmode{M_{\rm halo}}\else{$M_{\rm halo}$}\fi}}

\graphicspath{{./figs/}}

\begin{document}

\title{The BEHOMO project: $\Lambda$LTB $N$-body simulations}

\titlerunning{BEHOMO: $\Lambda$LTB $N$-body simulations}

\authorrunning{Marra et al.}

\author{V.~Marra\inst{\ref{oats},\ref{ifpu},\ref{cosmoufes}}\thanks{\email{valerio.marra@me.com}}
\and
T.~Castro\inst{\ref{oats},\ref{ifpu},\ref{infnTS}}
\and
D.~Camarena\inst{\ref{ppgcosmo}}
\and
S.~Borgani\inst{\ref{oats},\ref{ifpu},\ref{infnTS},\ref{units}}
\and
A.~Ragagnin\inst{\ref{unibo},\ref{oats},\ref{ifpu}}
}

\institute{
INAF -- Osservatorio Astronomico di Trieste, 34131, Trieste, Italy \label{oats}
\and
IFPU -- Institute for Fundamental Physics of the Universe, 34151, Trieste, Italy \label{ifpu}
\and
Núcleo Cosmo-ufes \& Departamento de Física, Universidade Federal do Espírito Santo, 29075-910, Vitória, ES, Brazil \label{cosmoufes}
\and
INFN -- Sezione di Trieste, 34100, Trieste, Italy \label{infnTS}
\and
PPGCosmo, Universidade Federal do Espírito Santo, 29075-910, Vitória, ES, Brazil \label{ppgcosmo}
\and
Dipartimento di Fisica, Sezione di Astronomia, Università di Trieste, 34143, Trieste, Italy \label{units}
\and
Dipartimento di Fisica e Astronomia ``Augusto Righi'', Alma Mater Studiorum Università di Bologna, 40129, Bologna, Italy \label{unibo}
}

\date{Received \today\ / Accepted \today}

\abstract
{Our Universe may feature large-scale inhomogeneities and anisotropies which cannot be explained by the standard model of cosmology, that is, the homogeneous and isotropic FLRW metric, on which the $\Lambda$CDM model is built, may not describe accurately observations.
Currently, there is not a satisfactory understanding of the evolution of the large-scale structure on an inhomogeneous background.}
{We start the cosmology {\bf be}yond {\bf homo}geneity and isotropy (BEHOMO) project and study the inhomogeneous $\Lambda$LTB model with the methods of numerical cosmology.
Understanding the evolution of the large-scale structure is a necessary step  to constrain inhomogeneous models with present and future observables and place the standard model on more solid grounds.
}
{We perform Newtonian $N$-body simulations, whose accuracy in describing the background evolution is checked against the general relativistic solution. The large-scale structure of the corresponding $\Lambda$CDM simulation is also validated.}
{We obtain the first set of simulations of the $\Lambda$LTB model ever produced. The data products consist of 11 snapshots between redshift 0 and 3.7 for each of the 68 simulations that have been performed, together with halo catalogs and lens planes relative to 21 snapshots, between redshift 0 and 4.2, for a total of approximately 180 TB of data.}
{We plan to study the growth of perturbations at the linear and nonlinear level, gravitational lensing, cluster abundances and proprieties. Data can be obtained upon request.
Further information is available at \href{https://valerio-marra.github.io/BEHOMO-project}{valerio-marra.github.io/BEHOMO-project}.}


\maketitle


\section{Introduction}

Several anomalous signals in cosmological observables have been emerging since the establishment of the $\Lambda$CDM  {(lambda Cold Dark Matter)} model as the standard model of cosmology more than two decades ago.
Particularly relevant here are the Hubble crisis, the  {Cosmic-Microwave-Background (CMB)} anomalies and the cosmic dipoles  {and bulk flows} \citep[see][and references therein]{Perivolaropoulos:2021jda}.
Such signals indicate anomalies that challenge the $\Lambda$CDM model and its foundations.
One may then ask if the Universe features large-scale inhomogeneities and anisotropies which cannot be explained by the standard paradigm, or, equivalently, if the homogeneous and isotropic   {Friedmann–Lemaître–Robertson–Walker (FLRW)} metric, on which the $\Lambda$CDM model is built, does not accurately describe  observations.
This constitutes the motivation for studying the Universe without assuming homogeneity and isotropy, trying instead to reconstruct the metric directly from observations \citep{Stebbins:2012vw}.

According to the standard reasoning, the validity of the FLRW metric is a  consequence of the observed isotropy of the Universe and the Copernican principle, which states that humans are not special observers.
Here, however, we are not advocating that the Universe is inhomogeneous and humans are special, rather that the scale at which there is homogeneity and isotropy could be larger than the  {commonly thought~$\approx$100~Mpc \citep[][]{Scrimgeour:2012wt,Laurent:2016eqo,Ntelis:2017nrj}, that is, the Cosmological principle may be valid at grander scales \citep[see][Section 8]{Abdalla:2022yfr}.}
Note that this scenario is not necessarily at odds with the observed approximate isotropy of the CMB, see the discussion of the Ehlers-Geren-Sachs theorem in \citet{Rasanen:2009mg}.

Inhomogeneous cosmology is undeniably a challenging subject as it would require a considerable theoretical and numerical effort to study its phenomenology. The absence of an FLRW background makes it particularly difficult to study early universe physics and predict, for instance, the CMB power spectrum.
Therefore, in order to present a viable program, here we consider a subclass of inhomogeneous cosmologies.
The basic requirement is that, at early times, one recovers a near-FLRW metric so that the standard inflationary paradigm is maintained and the physics that leads to the CMB remains basically unchanged.
In other words, we are considering a standard cosmology endowed with a non-standard large-scale structure that is dominated by growing modes.
This requirement effectively imposes restrictions on the free functions that characterize inhomogeneous metrics, considerably simplifying both analysis and statistical inference.
We name these inhomogeneous models as ``early-FLRW cosmologies.''

Clearly, early-FLRW cosmologies are constrained by CMB observations. Indeed, as showed by \citet{Valkenburg:2011ty}, perturbations at the last scattering surface of present-day contrast  $\approx$0.1 and size $\approx$1 Gpc would produce temperature fluctuations of $\Delta T \!\!\approx\!\! 50 \mu $K on a scale of $\approx\!\! 5^\circ$.
Similarly, too strong structures along the line of sight at $z \lesssim 1$ would be detected via the  {integrated Sachs-Wolfe (ISW)} effect%
: a present-day contrast  $\approx$0.1 and size  $\approx$300 Mpc would produce temperature fluctuations of order $\Delta T \!\!\approx\!\! 20\text{--}30 \mu $K \citep[see, for instance,][]{Zibin:2014vaa,Nadathur:2014tfa}.
To put this figure in perspective, the famous cold spot of the CMB features $\Delta T \!\!\approx\!\! 70 \mu $K across a $5^\circ$ region \citep{Vielva:2010ng}. Therefore, early-FLRW cosmologies are, at most, mildly nonlinear large-scale perturbations of the FLRW metric.

The inhomogeneities of early-FLRW cosmologies may be regarded as a particular type of primordial non-Gaussianity. Their distinguishing features are non-standard amplitudes and phases.
Indeed, they are characterized by bulk flows and coherent perturbations in the energy content of the universe at arbitrarily large scales.
In other words, large-scale homogeneity and isotropy are violated by the phases of these extra modes so that observations depend on the position of the observer and the notion of an average FLRW observer ceases to be meaningful \citep{Kolb:2009rp}.
Specifically, large-scale inhomogeneities alter observations both because they affect photon geodesics and the observer's local spacetime is perturbed. Of course, this is true also within the $\Lambda$CDM model but there the size of this effect is constrained by the standard perturbation spectrum. For example, cosmic variance on local measurements of $H_0$ is expected to be at most 1\% within $\Lambda$CDM \citep[][]{Camarena:2018nbr}.

The background evolution of early-FLRW cosmologies, that is, the evolution neglecting standard primordial perturbations, can be studied via exact solutions of General Relativity.
If considering more general scenarios, one may use linear-perturbation theory or simulations via  {codes that use general relativistic (GR) perturbation theory such as \texttt{gevolution} \citep{Adamek:2016zes} and \texttt{CONCEPT} \citep{Dakin:2021ivb}.}
A general consequence of spatial gradients is the occurrence of background shear, that is, the fact that the universe expands in an anisotropic way.

The scenario becomes more involved once primordial perturbations are added to the inhomogeneous background. First, there is the issue of the backreaction of small-scale perturbations on the average dynamics of the (possibly inhomogeneous) universe. Here, we assume that backreaction gives a negligible effect,  {as tested via GR simulations \citep[][]{Giblin:2015vwq,Bentivegna:2015flc,Adamek:2017mzb,Macpherson:2018btl}}. See Section~\ref{backreaction} for an overview of the backreaction proposal.
Second, because of spatial gradients, the standard primordial perturbations are coupled at first order so that standard perturbation theory does not hold in an inhomogeneous background.
Within the spherical  {Lemaître–Tolman–Bondi (LTB)} spacetime, this issue has been tackled by \citet{Zibin:2008vj,Clarkson:2009sc,Dunsby:2010ts,February:2013qza,Meyer:2014qla} via the numerical integration of the system of coupled equations and by  \citet{Nishikawa:2012we} via second-order perturbation theory.
It was concluded that the effect of spatial gradients could have an impact on the growth of perturbations.
However, a full perturbation theory and modeling is missing, hampering comparison with perturbation observables---the focus of current and next-generation surveys such as DES \citep{DES:2021wwk},\footnote{\href{https://www.darkenergysurvey.org}{www.darkenergysurvey.org}} DESI \citep{DESI:2016fyo},\footnote{\href{https://www.desi.lbl.gov}{www.desi.lbl.gov}} J-PAS \citep{Bonoli:2020ciz},\footnote{\href{http://www.j-pas.org}{www.j-pas.org}} LSST \citep{LSSTDarkEnergyScience:2012kar},\footnote{\href{https://www.lsst.org}{www.lsst.org}} Euclid \citep{Amendola:2016saw}\footnote{\href{https://www.euclid-ec.org}{www.euclid-ec.org}} and SKA \citep{Braun:2015zta}.\footnote{\href{https://www.skatelescope.org}{www.skatelescope.org}}

Here, we start the cosmology {\bf be}yond {\bf homo}geneity and isotropy (BEHOMO) project.
We propose a program that aims at addressing the modeling of linear and nonlinear perturbations and understanding the rich phenomenology of early-FLRW cosmologies. In order to do so, the basic idea is to apply the methods of numerical cosmology, as pioneered by \citet{Alonso:2010zv,Alonso:2012ds}.
The ultimate goal is to confront arbitrarily early-FLRW  inhomogeneous models with data from next-generation surveys.
The idea is to adopt Newtonian $N$-body simulations, whose accuracy in describing the background evolution shall be checked with general relativistic codes.
The basic methodology is to feed state-of-the-art $N$-body codes such as \texttt{GADGET} \citep[][]{Springel:2020plp} with special early-FLRW initial conditions so that early-FLRW cosmologies can reach the same resolution of standard $\Lambda$CDM simulations in approximately the same CPU time  {(except for the most nonlinear cases).}
This program will place the field of inhomogeneous cosmologies into the era of precision cosmology, on par with the $\Lambda$CDM model.

In this paper, we present the first suite of simulations for the simplest possible early-FLRW cosmologies: the spherically symmetric $\Lambda$LTB models.
We use the Lemaître-Tolman-Bondi (LTB) metric to model a spherical inhomogeneity on top of the standard $\Lambda$CDM model.
Though still a toy model, on a first approximation, one may regard the spatial gradients of the $\Lambda$LTB model as an archetype for more realistic structures with background shear.
We consider a set of high-resolution simulations with varying inhomogeneity size and depth, the two main physical parameters describing such a structure.
This will allow us to understand and model the effect of spatial gradients on the evolution of perturbations, which is necessary to confront inhomogeneous cosmologies with perturbation observables such as  {redshift-space distortions}, weak lensing and cluster abundances.
As said earlier, the grand goal is to study and then constrain the phenomenology of these beyond-$\Lambda$CDM inhomogeneities with observations \citep{Valkenburg:2012td,Redlich:2014gga,Camarena:2021mjr}.

In this presentation paper we will review the $\Lambda$LTB model in Section~\ref{LTBmodel}, discuss the numerical details of the inhomogeneous $N$-body simulations and their data products in Section~\ref{Nsim}, present the results of the simulations in Section~\ref{results}, and discuss the roadmap of the BEHOMO project%
\footnote{\href{https://valerio-marra.github.io/BEHOMO-project}{valerio-marra.github.io/BEHOMO-project}}
in Section~\ref{conclusions}.
Notation: we use `LTB metric' as opposed to `FLRW metric' but  `$\Lambda$LTB model' as opposed to  `$\Lambda$CDM model';  quantities without explicit radial dependence are relative to the FLRW background if pertinent; bold denotes vectors; $c=1$ is assumed unless stated otherwise.

\section{The $\Lambda$LTB model} \label{LTBmodel}

We consider early-FLRW $\Lambda$LTB models, that is, the $\Lambda$CDM model endowed with a spherical inhomogeneity, which is described via the exact LTB solution of Einstein's equations.
As we are considering early-FLRW cosmologies, this model is fully specified by the  radial profile function, whose basic parameters are the effective radius and depth of the inhomogeneity.

In this Section, after reviewing the formalism and dynamics of the LTB metric, we connect with the more standard Newtonianly perturbed FLRW metric and discuss the historical relevance of LTB models, putting coherently together results from many different papers.

\subsection{Metric}

In the comoving and synchronous gauge, the spherically symmetric LTB metric can be written as:
\begin{align} \label{metric}
\d s^2 = -\d t^2 + \frac{\apar^2(t,r)}{1-k(r)r^2}\d r^2 + \aperp^2(t,r)r^2 \, \d\Omega^2 \,,
\end{align}
where the longitudinal ($\apar$) and perpendicular ($\aperp$) scale factors are related by
$\apar = (\aperp r)'$, and a prime denotes partial derivation with respect to the coordinate radius $r$.
We will also adopt  the alternative notation $Y(t, r)\equiv \aperp r$ so that $Y'\! \equiv\apar$.
In the limit $k\to$\,constant and $\aperp=\apar=a$, we recover the FLRW metric, but here $k(r)$ is a free function named the LTB curvature function.

The two scale factors define two different Hubble rates:
\begin{align}
\Hperp(t,r) &\equiv \frac{\dot a_\perp}{\aperp} = \frac{\dot Y}{Y} \,, \\
\Hpar(t,r) &\equiv \frac{\dot a_{\|}}{\apar} = \frac{\dot Y'}{Y'} \,,
\end{align}
where a dot denotes partial derivation with respect to the coordinate time $t$.
This has important implications when confronting these models with observations.
 {For example, cosmic chronometers probe $\d z/\d t$ and so $\Hpar$, see Eq.~\eqref{eq:geotz}, radial baryon acoustic oscillations (BAO) also probe $\Hpar$, but angular BAO and supernovae probe the angular and luminosity distance, respectively, and so $\aperp$ and $\Hpar$, see Eqs.~(\ref{eq:geotz}-\ref{eq:ltbdist}). Combining these observables can then place interesting constraints on the background shear \citep{GarciaBellido:2008yq}:}
\begin{align}
\Sigma(t,r) = \frac{2}{3} \Big [  \Hpar(t,r) - \Hperp(t,r)   \Big ] \,.
\end{align}
As said earlier, the spatial gradient of the $\Lambda$LTB model is an archetype for more realistic structures.

\subsection{Dynamics}

By solving Einstein's equations for an irrotational dust source in the presence of a cosmological constant $\Lambda$, one obtains the equivalent of the Friedmann equation, which can be written as \citep[][Appendix B]{Enqvist:2007vb,Marra:2011zp}:
\begin{align} \label{fred}
\Hperp^2(t,r) &= {8 \pi G \over 3}   \rho_m^{\rm e}(t,r)  +  {8 \pi G \over 3} \rho_\Lambda-  { k(r) \over \aperp^2(t,r)} \,, 
\end{align}
where $\rho_\Lambda = \Lambda / 8 \pi G$, and the last term is the Euclidean average of the spatial Ricci scalar (the trace of the Ricci tensor of the spatial metric on the hypersurface of constant $t$):
\begin{align}
\frac{\cal{R}}{2} &= \frac{(k\,  r^2 Y)'}{Y^2 Y'}= \frac{k}{\aperp^2} + 2 \frac{k}{\aperp \apar} +\frac{k' r}{\aperp \apar} \,, \\
 {{\cal{R}}^{\rm e} \over 6} &= \frac{1}{6} \frac{ \int_{0}^{r}  {\cal{R}}  \, \d V_{\rm e}}{V_{\rm e}}  = { k(r) \over \aperp^2}    \xrightarrow{\text{FLRW}} \frac{k=\text{const}}{a^2} \,,
\end{align}
where the Euclidean volume element -- obtained by setting $k=0$ in eq.~\eqref{metric} --  is used:
\begin{align}
V_{\rm e}(t,r) &= \int_{0}^{r}    \d V_{\rm e}  =4 \pi \int_{0}^{r}  Y^{2} Y' \d \hat r = {4\pi \over 3} Y^{3} \,.
\end{align}
 {The fact that a Euclidean rather than proper average is used leads to backreaction, as discussed in Section~\ref{back}.}
Similarly, eq.~\eqref{fred} features the Euclidean average of the local matter density $\rho_m$:
\begin{align}
\rho_m (t,r)& =\frac{F'( r)}{4\pi Y^2(t, r) Y'(t, r)} \,, \label{rho} \\
F(r) &=   \int_{0}^{r}   \rho_m(t, r) \, \d V_{\rm e} \,,\\
\rho_m^{\rm e}(t,r) &= \frac{F(r)}{V_{\rm e}} \xrightarrow{\text{FLRW}} \rho_m (t)   \,, 
\end{align}
where the 
LTB mass function $F(r)$, a constant of integration, is another free function that gives the total gravitating mass up to the shell of coordinate radius $r$. The local density $\rho_m$ satisfies the continuity equation $\dot \rho_m + \theta \, \rho_m =0$, where  $\theta = \Hpar+ 2 \Hperp$ is the expansion scalar.
%
%
Note that, as the source is pressureless dust, without pressure gradients, both $F(r)$ and $k(r)$ do not depend on $t$. See \citet{Yamamoto:2015etj} for the case of the Lemaître metric with pressure.

Similarly to FLRW, one may interpret the curvature function as related to the total energy per unit of mass of the shell at coordinate radius $r$:
\begin{align} \label{energy}
E(r)\equiv - \frac{k \, r^2}{2} = \frac{1}{2} \dot Y^2(t,r) - \frac{G F(r)}{Y(t,r)} - \frac{1}{6}\Lambda Y^2(t,r) \,, 
\end{align}
where the first term of the energy function $E$ is the  kinetic energy per unit of mass of the shell $r$, the second term is the potential energy per unit of mass due to the total gravitating mass up to the shell $r$, and the third term is the usual contribution from the cosmological constant (as in the de Sitter-Schwarzschild metric). Note that, thanks to spherical symmetry, one is able to define a potential energy also in cases far away from nearly Newtonian ones and that the potential energy is related to the curvature~\citep[][]{Bondi:1947fta}.

 {Next, similarly to FLRW, one} can  rewrite eq.~\eqref{fred} using the equivalent of the density parameters in FLRW:
\begin{align} \label{fred2}
\frac{\Hperp^2(t,r)}{\Hperpo^2(r)} =\Omega_{m0}(r) \frac{\aperpo^3}{\aperp^3} + \Omega_{\Lambda0}(r)
+\Omega_{k0}(r) \frac{\aperpo^2}{\aperp^2} 
\end{align}
where the subscript 0 denotes a quantity evaluated at the present time $t_0$, and
\begin{align}
\Omega_{m0}(r) &=  \frac{ 2 G F(r)}{r^3 \aperpo^3 \Hperpo^2} & \Omega_{m} (t,r) &= \Omega_{m0}(r) \,  \frac{\Hperpo^2}{\Hperp^2} \frac{\aperpo^3}{\aperp^3} \,,  \\
\Omega_{\Lambda0}(r) &= \frac{\Lambda}{3 \Hperpo^2}  &  \Omega_{\Lambda} (t,r) &=\Omega_{\Lambda0} (r)\,  \frac{\Hperpo^2}{\Hperp^2} \,, \\
\Omega_{k0}(r) &= - \frac{ k(r)}{\aperpo^2  \Hperpo^2} & \Omega_{k} (t,r) &=\Omega_{k0}(r) \,  \frac{\Hperpo^2}{\Hperp^2}\frac{\aperpo^2}{\aperp^2} \,,
\end{align}
which satisfy $\Omega_m(t,r)  + \Omega_\Lambda(t,r) +\Omega_k(t,r)  =1$.

\subsection{Free functions and gauge fixing} \label{rgauge}

Eq.~\eqref{fred2} can be used to determine the age of the universe at a radial coordinate~$r$:
\begin{equation} \label{age}
   t-t_{bb}(r)\!=\!\frac{1}{\Hperpo(r)} \!\! \int^{\aperp(t,r) \over \aperpo(r)}_0 \!\!\!\!\!\!\!\! \frac{\d x}{\sqrt{\Omega_{m0}(r)/x\!+\!\Omega_{\Lambda 0}(r)x^2\!+\! \Omega_{k0}(r)}} ,
\end{equation}
where the big bang function $t_{bb}(r)$ is another arbitrary function, which sets the time since the big bang ($\aperp =0$).
If it were $t'_{bb}(r)\neq 0$, the initial singularity would have happened at different times for different shells so that  large inhomogeneities would develop in the past, as can be seen from eq.~\eqref{rho} with $Y \to 0$.
This clearly signals the presence of decaying modes, which would be strongly in contradiction with the inflationary paradigm and are excluded by the choice of a simultaneous big bang \citep{1977A&A....59...53S,Biswas:2006ub,Zibin:2008vj}.

Summarizing, we have seen that the LTB inhomogeneity is specified by three arbitrary functions, $F(r)$, $k(r)$ and $t_{bb}(r)$, which are related, together with $\aperpo$, by eq.~\eqref{age} so that one is not independent. Moreover, one can always make a redefinition of the radial coordinate.
Common gauge fixing are $F(r) \propto r^3$ or $\aperpo =$ constant.
It is then clear that one can choose $t_{bb}(r)$ and $k(r)$ as the free functions that specify the model.

Each gauge fixing has pros and cons. For example, $F(r) \propto r^3$ excludes the possibility that there is pure vacuum in some radial interval, and the moment of shell crossing -- the time at which $Y'=0$ so that $g_{rr}=0$ -- clearly depends on the gauge adopted.
The numerical codes that we use, \texttt{VoidDistances2020} \citep{Valkenburg:2011tm} and \texttt{FalconIC} \citep{Valkenburg:2015dsa},
adopt the choice $F(r) = 4 \pi M_0^4 r^3/3$, where $M_0$ is an arbitrary mass scale.

\subsection{Compensated inhomogeneity profile}

As discussed earlier, we will consider early-FLRW cosmologies in agreement with the standard scenario of inflation and, therefore, we will set:
\begin{align}
t_{bb}(r) =0 \,.
\end{align}

We are then left with the curvature function.
Here, we consider the case of an LTB inhomogeneity that matches exactly with the FLRW metric at the finite radius $r_b$ and not only asymptotically.
This simplified approach is convenient for the purposes of this work because it allows us to robustly simulate the LTB inhomogeneity inside of a bigger FLRW box.
The curvature function is modeled according to the monotonic profile:
\begin{align} \label{profi1}
k(r)= k_{b} + (k_c - k_{b}) \; W_3 (r/ r_b ) \,,
\end{align}
where $r_b$ is the coordinate radius of the spherical inhomogeneity,  {$k_b$ and $k_c$ are the curvature outside and at the center of the inhomogeneity, respectively,} and
$W_3$ is the function
\begin{align} \label{Pnf}
W_{n}(x)= \left\{\begin{array}{ll}
e^{- x^n / \left(1-x \right) } & \quad \mbox{ for } \quad   0  \le x < 1\\
0 & \quad  \mbox{ for }  \quad \phantom{0. \le} x \geq 1
\end{array}\right. \,.
\end{align}
The function $W_n(x)$ interpolates from $1$ to $0$
when $x$ varies from $0$ to $1$ while remaining differentiable, which implies that
that $k(r)$ is $C^\infty$ everywhere.
It is $\d^m W_n/\d x^m|_0=0$ for $0<m < n$, so that there is no cusp at the center.
In the limit $n\rightarrow \infty$, $W_n(x)$ approaches the tophat function.

For $r \ge r_b$ the curvature profile equals the curvature $k_{b}$ of the background FLRW such that for $r \ge r_b$ one exactly recovers the background $\Lambda$CDM model: $\aperp=\apar=a$. We can then define the local density contrast according to:
\begin{align}
\delta (t,r) =  \frac{\rho_m (t,r)}{ \rho_m (t)} - 1 \,,
\end{align}
and the (integrated) mass density contrast according to:
\begin{align}\label{eq:deltar}
\Delta (t,r) = \frac{\int_0^r  \delta(t,\bar r) \, \d V_{\rm e}}{V_{\rm e}}   = \frac{\Omega_m(t,r) \, \Hperp^2(t,r)}{\Omega_{m}(t)\, H^2(t)} -1 \,, 
\end{align}
where we used the Euclidean average in agreement with Eq.~\eqref{fred}.
Note that $\Delta(t,r=0)=\delta (t,r=0)$. We denote with $\delta_0$ the central contrast today, which is directly related to~$k_c$  {(see Eq.~\eqref{delta0lin} for the linear relation at early times).}

Note also that, because of the matching, it is by construction $\Delta(t,r=r_b)=\delta  (t,r=r_b)=0$.
This implies that the central under- or over-density at $0 \le r < r_t$, determined by the curvature $k_c$ at the center, is automatically compensated by a surrounding over- or under-dense shell at $r_t \le r < r_b$, where $r_t$ is the transition radius at which $\delta =0$.
A compensating over/underdense region is an expected feature of the standard large-scale structure: voids are surrounded by sheets and filaments, and superclusters by voids.
Note that it is $r_t=r_t(t)$, as in eq.~\eqref{rho} the volume element at the denominator is time dependent.

\subsection{Physical and lightcone distances}

The comoving radial coordinate $r$, because of the freedom in redefining it, does not possess physical meaning.
On the other hand, the proper distance between $r_1$ and $r_2$ ($\d t^2= \d \Omega^2 =0$ in eq.~\eqref{metric}) is:
\begin{align} \label{proper}
d_P =  \int_{r_1}^{r_2} \frac{Y'(t,r)}{\sqrt{1-k(r)r^2}}\d r
\simeq Y(t,r_2)-Y(t,r_1) \,, 
\end{align}
where the approximation holds for
\begin{align}
E \sim k(r)r^2 = \frac{Y^2}{\aperp^2/k} = \left (  \frac{Y}{\text{curv.~radius}}  \right )^2 \ll 1 \,.
\end{align}
Inside the inhomogeneity ($r<r_b$) the curvature radius is $\approx \aperp/\sqrt{k_c}$, while outside the LTB patch it is $a/ \sqrt{k_b}$.
We will consider models with $k_b=0$ so that the corrections to eq.~\eqref{proper} will be due only to the inhomogeneity.
We will see that these corrections are negligible also for Gpc-scale inhomogeneities ($E\ll 1$, see Fig.~\ref{example-t0}).

Using eq.~\eqref{proper} we can then define the corresponding FLRW comoving coordinate as:
\begin{align} \label{rout}
\chi= \frac{d_P}{a(t)} \overset{E \ll 1}{=} \frac{Y(t,r)}{a(t)} \,,
\end{align}
so that the FLRW and LTB physical distances coincide (note that $Y'\neq 0$). Thanks to the adopted matching condition, it is $\chi = r$ for $r\ge r_b$.
The coordinate $\chi$ is the one used in the numerical simulations.

Observationally, the time $t$ and radius $r$ as a function of the redshift $z$ are determined on the past lightcone of the central observer by the differential equations for radial null geodesics \citep[see, e.g.,][]{Chung:2006xh,Enqvist:2007vb}:
\begin{align}
\frac{\d t}{\d z} &= -\frac{1}{(1+z)\Hpar}\,, \label{eq:geotz}  \\
\frac{\d r}{\d z} &= \frac{\sqrt{1-k r^2}}{(1+z)\apar\Hpar} \, , \label{eq:geodesics}
\end{align}
with the initial conditions $t(0) = t_0$ and $r(0)=0$.
The area ($d_A$) and luminosity ($d_L$) distances are given~by
\begin{align}
d_A(z)&=a_\perp \big(t(z),r(z) \big) \, r(z) \,, \label{eq:ltbdistA} \\
d_L (z)&=(1+z)^2 d_A(z) \,.  \label{eq:ltbdist}
\end{align}

\begin{table}
\begin{center}
\setlength{\tabcolsep}{15pt}
\renewcommand{\arraystretch}{1.15}
\begin{tabular}{lc}
\hline
\hline
FLRW parameters & value \\
\hline
$H_0$ & 68 km/s/Mpc \\
$\Omega_m$ &  0.3 \\
$\Omega_k$ &  0  \\
\hline
\hline
Perturbation parameters & value \\
\hline
$\Omega_b$ & 0.048  \\
$\ln (10^{10}A_s)$ & 3.0  \\
$n_s$ &  0.97 \\
$\tau$ & 0.094  \\ 
$Y_{P}$ & 0.25  \\
$N_{\rm eff}$ & 3.046  \\
$\sum m_\nu$ & 0 \\
\hline
\hline
LTB parameters & value \\
\hline
$\delta_0$ & [-0.6, 0.6]  \\
$r_b$ &  [500, 4000] Mpc/$h$  \\
\hline
\hline
\end{tabular}
\caption{Parameters specifying the $\Lambda$LTB model.
The non-LTB parameters define the fiducial BEHOMO cosmology.
The amplitude $A_s$ of scalar perturbations and their spectral index $n_s$ are relative to the pivot scale $k_p=0.05$/Mpc.
This $\Lambda$CDM cosmology gives $\sigma_8 = 0.79364$ and $t_0=13.862$ Gyr.\protect\footnotemark
}
\label{tab:LLTB-pars}
\end{center}
\end{table}
\footnotetext{Radiation has been neglected as both the simulation and the GR calculations do not include radiation.}

\subsection{Model parameters}

The $\Lambda$LTB model is specified by the usual background FLRW parameters, that is, the Hubble constant $H_0$, the total matter density parameter $\Omega_m$ and the curvature parameter $\Omega_k$, by the standard perturbation parameters, that is, the baryon density parameter $\Omega_b$, the optical depth $\tau$, the helium fraction $Y_{P}$, effective number of relativistic species $N_{\rm eff}$, the total neutrino mass $\sum m_\nu$, the amplitude of the primordial power spectrum $A_s$ and its tilt $n_s$,
and, finally, by the LTB parameters, that is, the central curvature $k_c$ and the inhomogeneity radius $r_b$.
While numerically the profile is specified via the latter parameters, we will adopt, instead of $k_c$, the derived parameter $\delta_0$, which is the contrast today at the center of the inhomogeneity and is more intuitive to most cosmologists.
Table~\ref{tab:LLTB-pars} summarizes all the parameters and their fiducial values.

\subsection{Example of inhomogeneity} \label{example}

\begin{figure}
\centering 
\includegraphics[trim={.6cm .6cm .8cm .7cm}, clip, width=.89 \columnwidth]{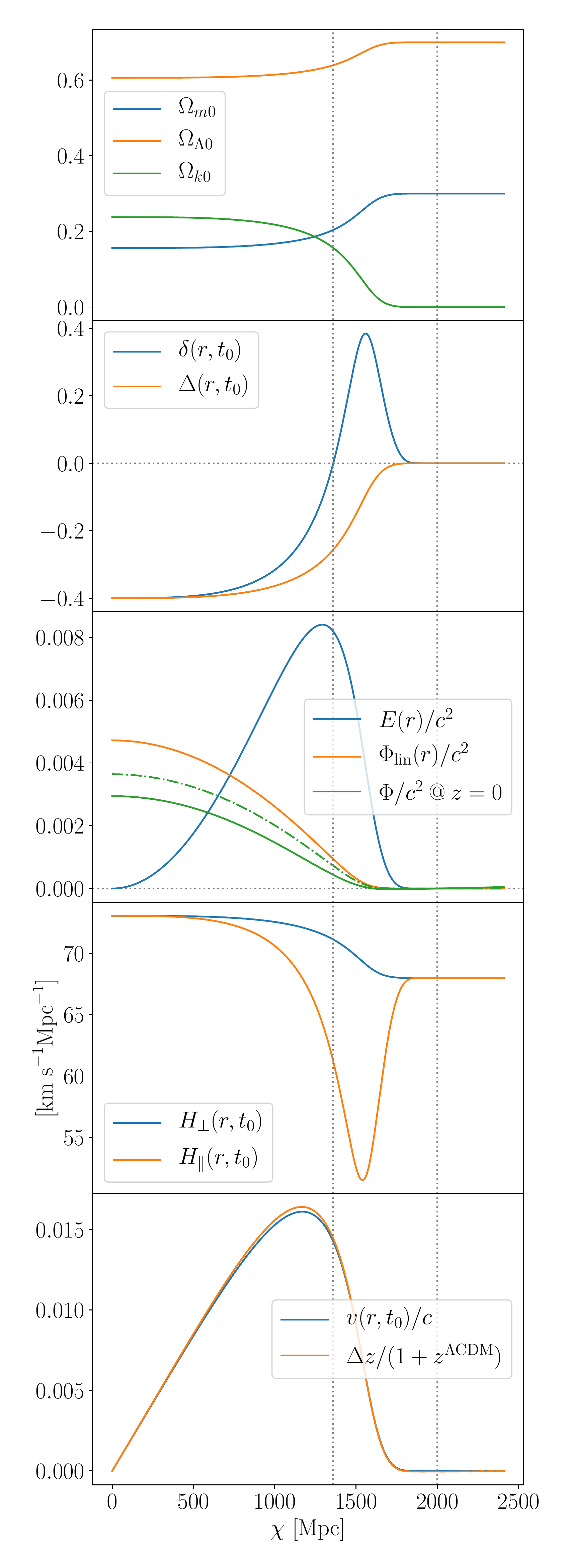}
\caption{LTB quantities as a function of the FLRW comoving coordinate $\chi$ at the present time $t_0$.
 {The two dotted lines mark the positions of the shells relative to $r_t$ and $r_b$.}
See Section~\ref{example}.
\label{example-t0}}
\end{figure}

\begin{figure}
\centering 
\includegraphics[trim={.6cm .8cm .8cm .8cm}, clip, width=1 \columnwidth]{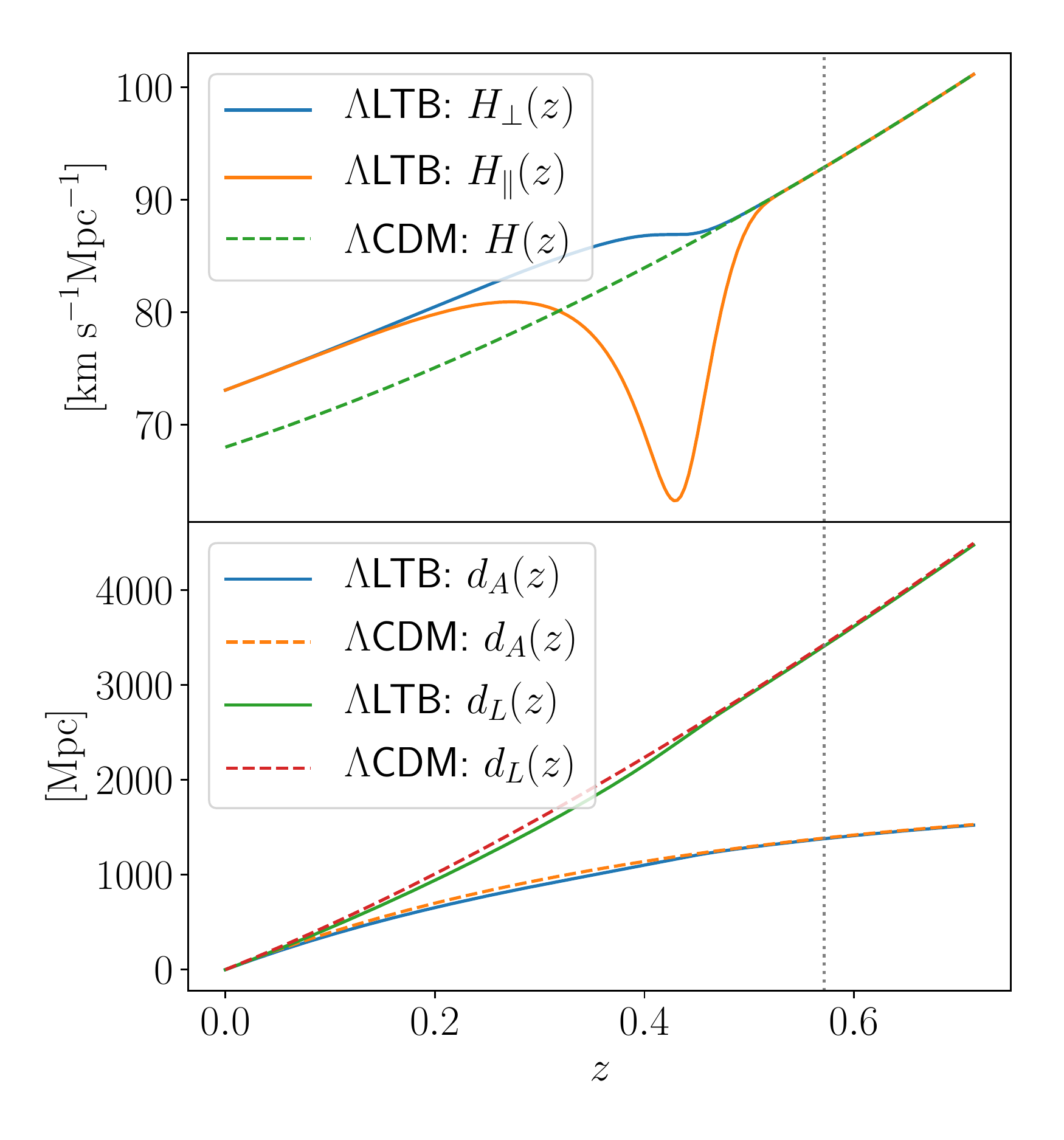}
\caption{LTB quantities as a function of redshift. See Section~\ref{example}.
\label{example-z}}
\end{figure}

Figure~\ref{example-t0} shows the relevant functions for the case of a central underdensity of present-day contrast $\delta_0=-0.4$ and comoving radius $r_b = 2000$ Mpc,  {and the fiducial BEHOMO cosmology of Table~\ref{tab:LLTB-pars}.} 
In particular, one can note that the interior of the inhomogeneity is an open FLRW universe (first panel from the top), that there is a compensating overdensity that surrounds the inner underdensity (second panel) and how the longitudinal Hubble rate deviates from the the perpendicular Hubble rate where there is a spatial gradient (fourth panel).
Also shown, for later use, are the linear and nonlinear Newtonian potentials together with the energy function (third panel), and the change in redshift induced by the inhomogeneity, together with the peculiar velocity defined in Eq.~\eqref{velo} (last panel).
Figure~\ref{example-z} shows the relevant functions on the lightcone as compared to their $\Lambda$CDM equivalent.

From Figure~\ref{example-t0} one can see that an inhomogeneity with a central underdensity of contrast $\delta_0=-0.4$ could solve the discrepancy  between local \citep{Riess:2021jrx} and high-redshift \citep{Planck:2018vyg} determinations of the Hubble constant: $H_0$ goes from the background value of 68 km/s/Mpc  to the local value of 73 km/s/Mpc.%
\footnote{
One can estimate the change in the expansion rate via linear perturbation theory. An adiabatic perturbation in density causes 
$\delta H_0/H_0 = - \frac{1}{3} f(\Omega_m)\delta \rho(t_0)/\rho(t_0)$,
where $f\simeq 0.5$ is the present-day growth rate for the concordance $\Lambda$CDM model.
}
This is the so-called local void scenario.
However, this scenario is ruled out by other observations.
\citet{Camarena:2021mjr,Camarena:2022iae} constrained the $\Lambda$LTB model using the latest available data from CMB, BAO, type Ia supernovae, local $H_0$, cosmic chronometers, Compton y-distortion and kinetic Sunyaev–Zeldovich  {(kSZ)} effect and showed that an underdensity around the observer as modeled within the $\Lambda$LTB model cannot solve the $H_0$ tension.
 {Appendix~\ref{ap:constraints} reports the latest constraints by \citet[]{Camarena:2021mjr} using the LTB parameters $\delta_0$ and $r_b$ that we adopt here.}

\subsection{Newtonianly perturbed FLRW metric}

One can regard the LTB inhomogeneity as a perturbation on top of the $\Lambda$CDM model. Here, we will connect the formalism of the previous sections with the one
of the Newtonianly perturbed FLRW metric:
%
\begin{align} \label{newton}
\d s^2 =    - \d \tilde t^2 (1+2 \Phi)  + a^2(\tilde t) (\d \tilde  r^2 +  \tilde  r^2 \d \Omega^2 ) (1-2\Phi)  \,,
\end{align}
where, for simplicity, we assumed a flat background FLRW metric.
This will be particular relevant as $N$-body simulations are in an FLRW background
\citep[see the discussion regarding the $N$-body gauge in][]{Fidler:2017pnb}.
This analysis will also be useful to highlight observational effects specific to $\Lambda$LTB inhomogeneities.
As we will see, in the case of sub-horizon inhomogeneities it is $\Phi \ll 1$.

By linearizing the LTB metric and considering a linear gauge transformation one finds that the Newtonian potential for $r< r_b$ is \citep{Biswas:2007gi,VanAcoleyen:2008cy}:
\begin{align} \label{linpot}
\Phi_{\rm lin}(r) & =  \frac{3}{5} \int_r^{r_b}  \frac{E(\bar r)}{\bar r} \, \d \bar r \sim  E \,, 
\end{align}
and $\Phi_{\rm lin}=0$ for $r \ge r_b$, where the potential is written as a function of the LTB coordinate.
 {Hereafter, the subscript ``lin'' refers to the fact that a linear gauge transformation is used; the potential is always linear, that is, a first-order perturbed quantity.}
Note also that $\Phi_{\rm lin}$ is constant in time, as should be for a linear matter perturbation in a matter-dominated universe. This description should be accurate at $z\gtrsim 10$.
The corresponding linear density contrast is:
\begin{align}
\Phi'_{\rm lin}(r) &= - \frac{3}{5} \frac{E( r)}{ r}\,, \label{phil-lin} \\
\nabla^{2}\Phi_{\rm lin}(r) &=\Phi''_{\rm lin} + 2 \frac{\Phi'_{\rm lin}}{r} = -\frac{3}{5} \left [  \frac{E'( r)}{ r}  + \frac{E( r)}{ r^2}    \right] \,, \\
\delta_{\rm lin}(t,r) &=  {\nabla^{2}\Phi_{\rm lin}(r) \over 4 \pi G   \rho_m(t) a(t)^{2} } \label{deltaLin}  \,,
\end{align}
where quantities without explicit radial dependence are relative to the FLRW background.
We took the derivative with respect to $r$ instead of the Newtonian gauge coordinate $\tilde r$, but the difference is second order.
Using Eq.~\eqref{deltaLin} together with Eqs.~\eqref{energy} and \eqref{profi1} one can find the initial evolution of the central density contrast as a function of the central curvature $k_c$:
\begin{align} \label{delta0lin}
\delta_{\rm lin}(t,0) = \frac{9 k_c}{40 \pi G    \rho_m(t) a(t)^{2} } \,.
\end{align}

One could use second-order perturbation theory to improve upon the latter linear description \citep[][]{Matarrese:1997ay}. However, given that, in general, the LTB inhomogeneity may feature nonlinear contrasts,%
\footnote{See \citet{Rigopoulos:2012xj} for an alternative approach that uses a gradient series expansion.}
we will now consider the potential as obtained via a nonlinear gauge transformation $\tilde t=\tilde t (t,r)$ and $\tilde r=\tilde r (t,r)$ which, following \citet{VanAcoleyen:2008cy}, is implicitly defined for $r< r_b$ by:
\begin{align}
Y(r,t)  &=  a(\tilde t) \tilde r \, \big (1-  \Phi(\tilde t,\tilde r) \big)  \,, \label{gauge1}\\
 t &= \tilde t+ a(\tilde t) \int_{\tilde r}^{ r_b}  v(\tilde t, \bar r) \, \d \bar r \,, \label{gauge2}
\end{align}
and by $\tilde r =r$ and $\tilde t = t$ for $r\ge r_b$,
%
%
where the peculiar velocity is:
\begin{align}
v (\tilde t, \tilde r) &= \dot Y(t,r) - \dot a (t) \, \tilde r=\dot Y(t,r)- H(t)\, Y(t,r) \nonumber \\
&= Y(t,r) \big[ \Hperp(t,r) - H(t)   \big] \,. \label{velo}
\end{align}
This gauge transformation will keep terms up to $\Phi,E \sim v^2$ and is valid for sub-horizon inhomogeneities.
From eq.~\eqref{gauge1} one sees that $\tilde r \xrightarrow{\Phi,E \ll 1} \chi $, that is, the coordinate $\chi$ defined in eq.~\eqref{rout} is indeed the one associated to the Newtonian gauge and, therefore, the one adopted by $N$-body simulations.

One can then use eqs.~(\ref{gauge1}-\ref{gauge2}) to change the LTB metric of eq.~\eqref{metric} into the Newtonian gauge of eq.~\eqref{newton} so as to find the the potential $\Phi$. Alternatively, one may proceed by inverting the Poisson equation:
\begin{align}
\nabla^{2}\Phi (\tilde r) = \frac{1}{\tilde r^2} \big (\tilde r^2 \Phi ' \big)' = 4 \pi G a^2 \left[ \frac{F'}{4 \pi Y^2 Y'}  \!-\! \frac{3 }{8 \pi G} \left(H^2- \frac{\Lambda}{3} \right)  \right ] ,
\end{align}
where the derivatives are with respect to the variable of the corresponding function. In particular, it is $\d \tilde r =  \frac{Y'}{a} (1+ \mathcal{O}(\Phi)) \d r$, so that one can integrate on $\tilde r$ and obtain:
\begin{align} \label{phil-nl}
\Phi ' = a\frac{G F}{Y^2}  - \frac{1}{2}a Y \left(H^2 -\frac{\Lambda}{3}\right)  \,,
\end{align}
where the constant of integration has been chosen in order to have $\Phi '(r_b) =0$ and the potential is expressed with respect to the LTB coordinate. Integrating again on $\tilde r$, one finally has:
\begin{align}
\Phi &= -\int_{r}^{r_b}  \frac{ Y' G F}{Y^2} \, \d \bar r  + \left ( H^2 - \frac{\Lambda}{3} \right) \left ( \frac{ Y_b^2}{4}-\frac{Y^2}{4} \right ) \label{phifull} \\
&=  \left ( H^2 - \frac{\Lambda}{3} \right) \left ( \frac{ Y_b^2}{4}-\frac{Y^2}{4} \right )+\frac{GF_b}{Y_b}-\frac{GF}{Y}  - \int_{r}^{r_b}  \frac{G F'}{Y} \, \d \bar r  \,, \nonumber
\end{align}
where $Y_b =a r_b $.
Note that $\Phi(r_b) =0$ and that we expressed the potential with respect to the LTB coordinate.
It is interesting to note $\nabla^{2}\Phi$ gives exactly the LTB contrast in LTB coordinates while the gauge transformation is only valid up to $\mathcal{O}(\Phi)$.
Figure~\ref{example-t0} (third panel, green solid curve) shows how the potential of Eq.~\eqref{phifull} decays during the cosmological-constant dominated phase as compared to the linear-gauge potential of Eq.~\eqref{linpot} during matter domination. Also shown (green dot-dashed line) is the linear perturbation result $\Phi(t, r) = \Phi_{\rm lin}(r) D(t)/ a(t)$, where $D$ is the $\Lambda$CDM growth function normalized at the matter-dominated epoch. The agreement with Eq.~\eqref{phifull} is perfect for linear LTB perturbations but it overestimates the value of the potential in the case of the nonlinear underdensity of Figure~\ref{example-t0}.

It is easy to verify that $\Phi '$ in Eq.~\eqref{phil-nl} reduces to the one of Eq.~\eqref{phil-lin} at early times:
\begin{align}
\tilde r \Phi' \overset{\eqref{phil-nl}}{=}& -  E + \frac{Y^2}{2} \left (  \Hperp^2 -H^2  \right )   \simeq - E + Y^2 H^2 \frac{\delta H}{H} \label{phili2}\\
=& -E + Y^2 H^2 \frac{\delta E}{5G F/Y}= -E \left ( 1 - \frac{Y^3 H^2}{5G F} \right )
= -\frac{3}{5}E \,, \nonumber
\end{align}
where we used Eq.~\eqref{energy} in the first equality, Eq.~\citep[31.14,][]{kaiserone} in the third equality and $F\simeq 4 \pi Y^3 \rho_m(t) /3$ in the last equality.
 {Moreover, using the result after the second equality, one has:
\begin{align}
v \overset{\eqref{velo}}{=}  Y \delta H \overset{\eqref{phili2}}{=}  \frac{\Phi '}{\dot a} + \frac{E}{\dot a r}
\overset{\eqref{phil-lin}}{=} - \frac{2}{3} \frac{\Phi '}{\dot a} \,,
\end{align}
in agreement with the expected matter-dominated result \citep[][Eq.~(18.1.9)]{coles2003cosmology}.}

\subsection{Observables in terms of the Newtonian potential}

If one uses the metric functions of the LTB metric of Eq.~\eqref{metric} then the effects of the inhomogeneities are exactly taken into account. However, 
it is important to discuss and review how the Newtonian potential affects observables. Indeed, the total potential will consist of the sum of the LTB potential and the potential relative to the primordial Gaussian perturbations, and the LTB potential may have observational effects in regimes in which the standard Gaussian potential is inconsequential.

A well-known result is that the redshift of photons are affected by perturbations according to \citep[see, for example,][]{Bonvin:2005ps}:
\begin{align}
\frac{\delta z}{1+z} \simeq ( \mathbf{v}_O  - \mathbf{v}_S) \cdot \mathbf{n} + (\Phi_O-\Phi_S) + 2a \! \int_{\chi_O}^{\chi_S}  \dot \Phi  \, \d \chi \, ,
\end{align}
where the vector $\mathbf{n}$ gives the direction of the source $S$ with velocity $\mathbf{v}_S$ with respect to the observer $O$ with velocity $\mathbf{v}_O$.
In the following, we will only consider the contribution from the LTB potential, but there are of course also contributions from standard-model perturbations.
The change in redshift can be interpreted as the sum of three effects.

First, there is the differential Doppler shift due to the peculiar motion of source and observer.
This contribution is zero if the observer and the source are placed outside the inhomogeneity or (one of them) at its center. Otherwise one expects a contribution that is proportional to $v \sim Y \Delta H \sim r_b/ r_{\rm hor}$ where $r_{\rm hor} = H^{-1}$ is the Hubble radius. This contribution is large for gigaparsec-scale inhomogeneities and  can significantly alter the luminosity distance-redshift relation and it was indeed used to fit supernova data without dark energy in the void scenario, see Section~\ref{void-scenario} for a historical note.
Figure~\ref{example-t0} (bottom panel) shows this effect for an underdensity of contrast $\delta_0=-0.4$ and comoving radius $r_b = 2000$ Mpc.
Also shown is the peculiar velocity as defined in Eq.~\eqref{velo}. One can see that most of the change in redshift can indeed be attributed to a Doppler shift.

The second term gives the so-called Sachs–Wolfe effect, that is, the differential gravitational redshift due to the gravitational potentials at the source's and observer's positions \citep{1967ApJ...147...73S}.
This contribution is zero if the observer and the source are placed outside the inhomogeneity. Otherwise, by comparing eq.~\eqref{energy} and~\eqref{velo} it is easy to see that $E \sim \dot Y^2/2-Y^2H^2/2 \sim - v^2/2$ so that the potential of eq.~\eqref{linpot} is quadratic in the velocities, $\Phi \propto v^2$.
It then follows that $\Phi \propto (r_b/ r_{\rm hor})^2$ and the Sachs–Wolfe effect is subdominant with respect to the Doppler shift.
Again from the analysis of Figure~\ref{example-t0} (bottom panel), one can see that $\delta z/1+z \simeq  -4 \times 10^{-5}$ at $r_b$, significantly smaller than the Doppler shift that occurs for source inside the inhomogeneity.

Finally, the last term is the integrated Sachs-Wolfe effect (ISW), which is present only if the (first-order) gravitational potential evolves with time and is responsible for a nontrivial correlation between CMB anisotropies and the large-scale structure.
At high redshift, $10 \lesssim z \lesssim 100$, the standard model is very close to the flat matter-dominated Einstein-de Sitter model.
It is then well-known that the (linear) potential is time-independent so that the ISW contribution is zero.
At later times, however, there are two contributions. First, the universe enters the cosmological constant-dominated phase: this is responsible for the (linear) ISW effect. Second, structures may enter the nonlinear regime so that the so-called Rees-Sciama (RS) effect cannot be neglected.
As discussed in \citet{Cai:2010hx}, the nonlinear RS correction to the ISW effect acts differently for over and underdensities.
\citet{Biswas:2007gi} explicitly showed that these contributions are suppressed according to $\propto (r_b/ r_{\rm hor})^3$.

For the mildly nonlinear large structures here considered, one expects that RS is subdominant with respect to ISW \citep{Sakai:2008fi}. In this case, the potential decays according to the linear ISW modeling:
\begin{align}
\dot \Phi  = \frac{3}{2} \Omega_{m0} H_0^2  G(z) P(r) \,,
\end{align}
where the (nonlinear) potential is obtained via Eq.~\eqref{phifull} and the ISW growth factor $G$ is:
\begin{align}
G(z)  =(1+z) H(z) [1-f(z)] D(z) \,,
\end{align}
where $D$ is the linear growth function, $f \equiv \d \ln D /\d \ln a \simeq \Omega_{m}^\gamma (t)$ with $\gamma=6/11 + 15/11^3 (1-\Omega_{m}(t))$ is the linear growth rate \citep{Wang:1998gt}, and $P(r)$ encodes the information on the inhomogeneity profile. See \citet{Nadathur:2011iu,Flender:2012wu} for a thorough discussion.


As thoroughly discussed in \citet{Hui:2005nm}, a perturbation in the redshift affects the luminosity distance. As we have seen, the LTB metric features possibly large contributions from peculiar velocities that are instead negligible in the standard paradigm. This means that proper care has to be adopted when analyzing these models on the lightcone.
The other important effect to consider is lensing, which modifies the observed flux of an object without changing its redshift. However, in this case, the total effect will be directly computed by binning mass in a suitable number of lens planes. Indeed, the total lensing effect is the sum of the contribution of the LTB potential with the one of the Gaussian perturbations' potential, and these two components make up the various lens planes.

\subsection{Scale invariance} \label{scale-inv}

As the dynamical equation \eqref{fred} does not present gradients, the dynamics of the LTB model is scale invariant.
This is due to spherical symmetry and the fact that the energy-momentum tensor is dust.
The former implies a vanishing magnetic Weyl tensor and consequently no gravitational waves; the latter implies no pressure and so no sound waves.
In other words, no direct communication can exist between neighboring worldlines and for this reason such spacetimes were dubbed `silent' \citep{1993PhRvD..47.1311M,1995ApJ...445..958B}.
In particular, pressure gradients would  transfer energy between shells and make the energy function $E$ and mass function $F$ time dependent~\citep[see][]{Marra:2011zp}.

Formally, starting from the solution of Eq.~\eqref{fred} for a given $r_b$, one can obtain a scaled inhomogeneity with coordinate $\hat r = \lambda r$ and size $\hat r_b =\lambda r_b$.
The Friedmann-like equation is then:
\begin{align}
\frac{\dot{ {\hat a}}_\perp(t,\hat r)}{\hat{a}_\perp(t,\hat r)} = {8 \pi G \over 3} \frac{M_0^4}{ \hat{a}_\perp^3(t,\hat r)} +  {8 \pi G \over 3}  \rho_\Lambda-  {\hat k(\hat r) \over \hat{a}_\perp^2(t,\hat r)} \,,
\end{align}
where we adopted the gauge fixing $F(r) = 4 \pi M_0^4 r^3/3$ and the functions relative to the scaled inhomogeneity are defined according to:
\begin{align}
{\hat a}_{\perp/\|}(t,\hat r) & = a_{\perp/\|}(t,\hat r/\lambda) \,, \\ 
{\hat H}_{\perp/\|}(t,\hat r) & = H_{\perp/\|}(t,\hat r/\lambda) \,, \\ 
{\hat k}(t,\hat r) & = k(t,\hat r/\lambda) \,, \\ 
{\hat \rho}_m(t,\hat r) & = \rho_m(t,\hat r/\lambda) \,, \\ 
{\hat Y}(t,\hat r) & = \lambda \, Y(t,\hat r/\lambda) \,, \\ 
{\hat v}(t,\hat r) & = \lambda \, v(t,\hat r/\lambda) \,, \\ 
{\hat E}(t,\hat r) & = \lambda^2 E(t,\hat r/\lambda) \,, \\ 
{\hat \Phi}(t,\hat r) & = \lambda^2 \Phi(t,\hat r/\lambda) \,, \\ 
{\hat F}(t,\hat r) & = \lambda^3 F(t,\hat r/\lambda) \,.
\end{align}
Starting from one numerical solution, one can then obtain a family of solutions by varying $\lambda$.

Note that velocities, and so Doppler effects, are proportional to $\lambda$, explaining why one needs a large inhomogeneity to sizably change the luminosity distance-redshift relation as in the void scenario discussed in see Section~\ref{void-scenario}.
Also, the energy function and the potential scale quadratically with the size so that one expects strong features in the power spectrum of large inhomogeneities.


\begin{table*}[p]
\tiny
\centering
\caption{BEHOMO suite of simulations. 
The first line of each Box series describes the corresponding $\Lambda$CDM simulation. Masses are defined according to 200m. The number of grid elements of the particle mesh is twice the number of particles per dimension.}
\label{tab:sims}
\setlength{\tabcolsep}{6pt}
\renewcommand{\arraystretch}{.9}
\begin{tabular}{l|lp{1cm}p{1cm}p{1.53cm}p{.9cm}lp{1.3cm}p{1.3cm}p{1.3cm}p{1.95cm}}
\hline
\hline
 & $\delta_0$ & $r_b$ (Gpc/$h$) & $L_{\rm box}$ (Gpc/$h$) & grav.\ soft.\ at $z=0$ (kpc/$h$)& $N_{\rm part}$     & $M_{\rm part}$ ($M_\odot/h$) & $M^{\rm min}_{\rm halo}$ ($M_\odot/h$) & $M^{\rm max}_{\rm halo}$ ($\textbf{}M_\odot/h$) & $N_{\rm halo}$   & {CPUh increase} (wrt $\Lambda$CDM sim)\\
\hline
\hline
Box 1 & --         & --   & 0.5 & 6.4  & $1024^3$    &  $9.7\times 10^9$     & $4.8\times 10^{11}$  & $1.7 \times 10^{15}$   &    $3.0 \times 10^{6}$   & -- \\ 
& \phantom{-}0.10         & 0.2   & " & "  & " &   "      & "   & $1.7 \times 10^{15}$ &   $3.0 \times 10^{6}$ & 1\%    \\
& \phantom{-}0.15         & "   & " & "  & " &   "    & "   & $1.6 \times 10^{15}$ &$3.1 \times 10^{6}$  &   0\%    \\
& \phantom{-}0.20         & "   & " & "  & " &   "     & "   & $1.6 \times 10^{15}$& $3.0 \times 10^{6}$  &   0\%    \\
& \phantom{-}0.30         & "   & " & "  & " &   "     & "   & $1.5 \times 10^{15}$ & $3.0 \times 10^{6}$   &  1\%    \\
& \phantom{-}0.45         & "   & " & "  & " &   "    & "   & $1.4 \times 10^{15}$ & $3.0 \times 10^{6}$  &   1\%    \\
& \phantom{-}0.60         & "   & " & "  & " &   "    & "   & $1.6 \times 10^{15}$ & $3.0 \times 10^{6}$  &  3\%     \\
& -0.10         & "   & " & "  & " &   "  &  "     & $1.8 \times 10^{15}$ &  $3.0 \times 10^{6}$  &   0\%   \\
& -0.15         & "   & " & "  & " &   "  &  "     &  $2.0 \times 10^{15}$ &$3.0 \times 10^{6}$&   0\%    \\
& -0.20         & "   & " & "  & " &   "  &  "      &  $2.2 \times 10^{15}$  & $3.0 \times 10^{6}$ &  1\%   \\
& -0.30         & "   & " & "  & " &   "  &  "      & $2.9 \times 10^{15}$ &$3.0 \times 10^{6}$  &  1\%     \\
& -0.45         & "   & " & "  & " &   "  &  "      & $3.4 \times 10^{15}$ &$3.0 \times 10^{6}$  &  4\%     \\
& -0.60         & "   & " & "  & " &   "  &  "      & $4.1 \times 10^{15}$  & $3.0 \times 10^{6}$  & 8\%    \\
\hline
Box 2 & --         & --   & 1.0 & 12.8  &  $1024^3$    &  $7.8\times 10^{10}$     & $3.9\times 10^{12}$    &     $2.5\times 10^{15}$ &   $3.7 \times 10^{6}$  & -- \\ 
& \phantom{-}0.10         & 0.4   & " & "  & " &   "      & "   &     $2.4 \times 10^{15}$ & $3.7 \times 10^{6}$   & 0\% \\
& \phantom{-}0.15         & "   & " & "  & " &   "     & "   &     $2.5 \times 10^{15}$ & $3.7 \times 10^{6}$   & 0\% \\
& \phantom{-}0.20         & "   & " & "  & " &   "    & "   &     $2.5 \times 10^{15}$ & $3.7 \times 10^{6}$   & 0\% \\
& \phantom{-}0.30         & "   & " & "  & " &   "     & "   &     $2.5 \times 10^{15}$& $3.7 \times 10^{6}$   & 5\% \\
& \phantom{-}0.45         & "   & " & "  & " &   "     & "   &     $2.7 \times 10^{15}$& $3.7 \times 10^{6}$   & 21\% \\
& \phantom{-}0.60         & "   & " & "  & " &   "    & "   &     $3.2 \times 10^{15}$ &$3.7 \times 10^{6}$    & 3\% \\
& -0.10         & "   & " & "  & " &   "  &  "    &     $2.4 \times 10^{15}$ & $3.7 \times 10^{6}$   & 2\% \\
& -0.15         & "   & " & "  & " &   "  &  "     &     $2.4 \times 10^{15}$ & $3.7 \times 10^{6}$    & 2\% \\
& -0.20         & "   & " & "  & " &   "  &  "     &     $2.5 \times 10^{15}$&$3.7 \times 10^{6}$    & 0\% \\
& -0.30         & "   & " & "  & " &   "  &  "     &     $2.5 \times 10^{15}$ & $3.7 \times 10^{6}$   & 22\% \\
& -0.45         & "   & " & "  & " &   "  &  "     &     $4.0 \times 10^{15}$ & $3.7 \times 10^{6}$   & 5\% \\
& -0.60         & "   & " & "  & " &   "  &  "     &     $5.6 \times 10^{15}$ &$3.7 \times 10^{6}$    & 16\% \\
\hline
Box 3 & --         & --   & 1.5 & 9.6   & $2048^3$    &  $3.3\times 10^{10}$     & $1.6\times 10^{12}$    &     $3.9 \times 10^{15}$ &  $2.8 \times 10^{7}$  & -- \\ 
& \phantom{-}0.10         & 0.6   & " & "  & " &   "       & "   &     $3.9 \times 10^{15}$ & $2.8 \times 10^{7}$  & 1\% \\
& \phantom{-}0.15         & "   & " & "  & " &   "     & "   &     $3.9 \times 10^{15}$ &  $2.8 \times 10^{7}$  & 0\% \\
& \phantom{-}0.20         & "   & " & "  & " &   "     & "   &     $3.8 \times 10^{15}$ & $2.8 \times 10^{7}$  & 0\%  \\
& \phantom{-}0.30         & "   & " & "  & " &   "     & "   &     $3.8 \times 10^{15}$& $2.8 \times 10^{7}$  & 0\% \\
& \phantom{-}0.45         & "   & " & "  & " &   "     & "   &     $3.8 \times 10^{15}$ & $2.8 \times 10^{7}$  & 5\% \\
& \phantom{-}0.60         & "   & " & "  & " &   "     & "   &     $5.9 \times 10^{15}$ &  $2.8 \times 10^{7}$  & 8\% \\
& -0.10         & "   & " & "  & " &   "  &  "     &     $3.9 \times 10^{15}$ & $2.8 \times 10^{7}$  & 1\% \\
& -0.15         & "   & " & "  & " &   "  &  "      &     $4.0 \times 10^{15}$&  $2.8 \times 10^{7}$ & 13\% \\
& -0.20         & "   & " & "  & " &   "  &  "      &     $4.0 \times 10^{15}$ & $2.8 \times 10^{7}$   & 0\% \\
& -0.30         & "   & " & "  & " &   "  &  "     &     $4.0 \times 10^{15}$ & $2.8 \times 10^{7}$   & 5\% \\
& -0.45         & "   & " & "  & " &   "  &  "      &     $4.0 \times 10^{15}$ & $2.8 \times 10^{7}$   & 30\%  \\
& -0.60         & "   & " & "  & " &   "  &  "      &     $8.2 \times 10^{15}$ &  $2.7 \times 10^{7}$  & 38\% \\
\hline
Box 4 & --         & --   & 2.0 & 12.8   & $2048^3$    &  $7.8\times 10^{10}$     & $3.9\times 10^{12}$    &     $4.3 \times 10^{15}$ &$3.0 \times 10^{7}$ &-- \\ 
& \phantom{-}0.10         & 0.8   & " & "  & " &   "       & "   &     $4.3 \times 10^{15}$& $3.0 \times 10^{7}$  & 0\% \\
& \phantom{-}0.15         & "   & " & "  & " &   "    & "   &     $4.2 \times 10^{15}$& $3.0 \times 10^{7}$  & 0\% \\
& \phantom{-}0.20         & "   & " & "  & " &   "     & "   &     $4.2 \times 10^{15}$& $3.0 \times 10^{7}$  & 0\% \\
& \phantom{-}0.30         & "   & " & "  & " &   "     & "   &     $4.3 \times 10^{15}$&$3.0 \times 10^{7}$   & 3\% \\
& \phantom{-}0.45         & "   & " & "  & " &   "    & "   &     $4.3 \times 10^{15}$& $3.0 \times 10^{7}$  & 3\% \\
& \phantom{-}0.60         & "   & " & "  & " &   "     & "   &     $4.5 \times 10^{15}$ &$3.0 \times 10^{7}$   & 6\% \\
& -0.10         & "   & " & "  & " &   "  &  "      &     $4.3 \times 10^{15}$& $3.0 \times 10^{7}$  & 0\% \\
& -0.15         & "   & " & "  & " &   "  &  "      &     $4.2 \times 10^{15}$& $3.0 \times 10^{7}$  & 0\% \\
& -0.20         & "   & " & "  & " &   "  &  "     &     $4.3 \times 10^{15}$& $3.0 \times 10^{7}$  & 3\% \\
& -0.30         & "   & " & "  & " &   "  &  "     &     $4.3 \times 10^{15}$&$3.0 \times 10^{7}$   & 5\% \\
& -0.45         & "   & " & "  & " &   "  &  "      &     $4.7 \times 10^{15}$& $3.0 \times 10^{7}$  & 30\% \\
& -0.60         & "   & " & "  & " &   "  &  "      &     $6.9 \times 10^{15}$& $2.9 \times 10^{7}$  & 41\% \\
\hline
Box 5 & --         & --   & 3.0 & 19.2   & $2048^3$    &  $2.6\times 10^{11}$     & $1.3\times 10^{13}$    &     $4.1 \times 10^{15}$&$3.2 \times 10^{7}$ &-- \\ 
& \phantom{-}0.10         & 1.2   & " & "  & " &   "       & "   &     $4.1 \times 10^{15}$& $3.2 \times 10^{7}$  & 1\% \\
& \phantom{-}0.15         & "   & " & "  & " &   "     & "   &     $4.1 \times 10^{15}$&$3.2 \times 10^{7}$   & 2\% \\
& \phantom{-}0.20         & "   & " & "  & " &   "     & "   &     $4.4 \times 10^{15}$& $3.2 \times 10^{7}$   & 6\% \\
& \phantom{-}0.30         & "   & " & "  & " &   "     & "   &     $4.9 \times 10^{15}$&$3.2 \times 10^{7}$   & 8\% \\
& \phantom{-}0.45         & "   & " & "  & " &   "     & "   &     $5.5 \times 10^{15}$&$3.2 \times 10^{7}$   & 10\% \\
& \phantom{-}0.60         & "   & " & "  & " &   "     & "   &     $5.9 \times 10^{15}$&$3.2 \times 10^{7}$   & 16\% \\
& -0.10         & "   & " & "  & " &   "    & "   &     $4.7 \times 10^{15}$ & $3.2 \times 10^{7}$   & 1\% \\
& -0.15         & "   & " & "  & " &   "     & "   &     $5.1 \times 10^{15}$ & $3.2 \times 10^{7}$   & 3\% \\
& -0.20         & "   & " & "  & " &   "     & "   &     $5.4 \times 10^{15}$ & $3.2 \times 10^{7}$  & 7\% \\
& -0.30         & "   & " & "  & " &   "     & "   &     $6.0 \times 10^{15}$ & $3.2 \times 10^{7}$   & 12\% \\
& -0.45         & "   & " & "  & " &   "     & "   &     $7.4 \times 10^{15}$ & $3.2 \times 10^{7}$   & 35\% \\
& -0.60         & "   & " & "  & " &   "     & "   &     $1.0 \times 10^{16}$ & $3.1 \times 10^{7}$  & 67\% \\
\hline
Box 6 & --  & --   & 4.0 & 22.7   & $2304^3$    &  $4.4\times 10^{11}$     & $2.2\times 10^{13}$    &     $4.9 \times 10^{15}$ & $4.5 \times 10^{7}$  & -- \\ 
& \phantom{-}0.20         & 1.6   & "   & " &   "  &  "   & "   &     $5.4 \times 10^{15}$ &$4.5 \times 10^{7}$   & 1\% \\
& -0.20         & "   & " & "  & "   &  "   & "   &     $5.0 \times 10^{15}$ & $4.5 \times 10^{7}$   & 1\% \\
\hline
\hline
\end{tabular}%
\end{table*}


\subsection{A historical note on LTB void models}
\label{void-scenario}

The LTB model has been studied extensively in the literature as an alternative to dark energy.
The relevant case was of an observer sitting near the center of a gigaparsec-scale underdensity. It is easy to understand how such an observer would see apparent acceleration: most of our cosmological observables are confined to the lightcone and, hence, temporal changes can be associated with spatial changes along photon geodesics.
The LTB void model then replaces ``faster expansion now than before'' with ``faster expansion here than there.''
Mathematically, the directional derivative on the past light cone follows $\d/\d t \approx \partial/\partial t - \partial/\partial r$ and the  accelerating expansion can be explained by  $H'(r) < 0$ \citep{Enqvist:2007vb}.
For 15 years the LTB model was phenomenologically viable, although suffered the extreme fine-tuning of the observer's position (see \citealt{Marra:2011ct,Bolejko:2011jc,Clarkson:2012bg} and references therein).
More importantly, it constituted perhaps the only example of a paradigm which departed abruptly from $\Lambda$CDM. 
This allowed cosmologists to ask new questions and develop new methodologies.

However, in 2011, two papers ruled out the LTB model, which already showed problems when confronted with more and more data~\citep{GarciaBellido:2008gd,Moss:2010jx,Biswas:2010xm}. \citet{Zhang:2010fa} \citep[see also][]{Moss:2011ze,Bull:2011wi} showed that void models without decaying modes produce a too large kSZ signal, and \citet{Zibin:2011ma} showed that void models with sizable decaying modes (which could possibly have a small kSZ signal) are ruled out because of y-distortion.

Despite the latter strong evidence against void models as alternatives to dark energy, one has to point out that those studies considered a homogeneous radiation field. In other words, inhomogeneities were present only in the matter component.
\citet{Clarkson:2010ej,Lim:2013rra} considered the more consistent scenario of inhomogeneities also in the radiation and showed that this could alter kSZ and y-distortion predictions.

\subsection{The backreaction proposal} \label{backreaction}

 {Because of the non-linear nature of General Relativity, the average of the solution of Einstein's equations for an inhomogeneous metric is not the solution of Einstein's equations for the average of the metric, that is, the operation of smoothing does not commute with solving Einstein's equations, $\langle G_{\mu \nu}(g_{\alpha \beta}) \rangle \neq G_{\mu \nu}({\langle g_{\alpha \beta} \rangle})$.}
Consequently, the Friedmann equation -- valid for a homogeneous universe -- features corrections in the form of extra sources \citep[see][which is considered the backreaction manifesto]{Ellis1984}.

In the early 2000's, right after the first analyses indicating the acceleration of the Universe's expansion by \citet{SupernovaSearchTeam:1998fmf,SupernovaCosmologyProject:1998vns}, it has then been asked if dark energy could actually be explained via the extra sources generated by the nonlinear smoothing, that is, via the backreaction of small-scale inhomogeneities into the large-scale dynamics of the universe.
This scenario would elegantly explain the biggest problem of dark energy: why does it appear at $z\approx 1$?
The answer would be that structures go nonlinear at $z\approx 1$, transforming a fine tuning into a prediction.
It is important to point out that, when the backreaction scenario was proposed, the equation of state $w$ of dark energy was poorly constrained.  {This is particularly relevant because the effective $w$ that one would measure, if dark energy were caused by backreaction, is not expected to be close to $-1$, the value relative to the cosmological constant. In other words, while the present-day tight constraint $w=-1.03 \pm 0.03$ \citep{DES:2021wwk} is, within the backreaction proposal, a coincidence, it is instead a necessary condition for the $\Lambda$CDM model.}


This proposal started a heated debate on the magnitude of the backreaction effect, which proved difficult to be estimated via (semi) analytical techniques. See \citet{Clarkson:2011zq,Buchert:2015iva,Green:2014aga}, and references therein.

In the past few years, the scientific consensus on the relevance of backreaction in cosmology has been sought via the methods of numerical relativity.
It seems that backreaction produces a negligible correction to the universe dynamics, although the methodology that has been adopted has some limitations.
On one hand fully general relativistic codes are used but the implementation of the fluid description of the matter sector raises questions regarding the modeling of the non linear structure formation which is dominated by halo mergers and shell crossing \citep{Giblin:2015vwq,Bentivegna:2015flc, Macpherson:2018btl}.
 {On the other hand,  particle-based modeling is adopted at the price of using the weak-field expansion of Einstein’s equations \citep{Adamek:2017mzb}.
However, codes that adopt a particle description alongside numerical relativity \citep[][]{East:2017qmk,Giblin:2018ndw,Daverio:2019gql}, including the recent code \texttt{GRAMSES} \citep{Barrera-Hinojosa:2019mzo,Barrera-Hinojosa:2020arz}, are set to provide important progress towards a definitive answer to the questions raised by the backreaction proposal \citep[][]{Adamek:2020jmr}.}

\begin{table}
\tiny
\centering
\caption{Snapshots that were saved during the runs for the generation of the halo catalog and lens planes. Only even-numbered snapshots are kept for long-term storage.}
\label{tab:snaps}
\setlength{\tabcolsep}{12pt}
\renewcommand{\arraystretch}{1.}
\begin{tabular}{cccc}
\hline
\hline
snapshot  & $a$ & $z$ &  comov.~dist.~(Mpc/$h$) \\
\hline
 0 & 0.02 & 49 & 8360 \\
 1 & 0.191 & 4.225 & 5125 \\
 2 & 0.212 & 3.715 & 4875 \\
 3 & 0.234 & 3.275 & 4625 \\
 4 & 0.257 & 2.892 & 4375 \\
 5 & 0.281 & 2.557 & 4125 \\
 6 & 0.307 & 2.261 & 3875 \\
 7 & 0.333 & 1.999 & 3625 \\
 8 & 0.362 & 1.765 & 3375 \\
 9 & 0.391 & 1.555 & 3125 \\
 10 & 0.423 & 1.366 & 2875 \\
 11 & 0.456 & 1.194 & 2625 \\
 12 & 0.491 & 1.037 & 2375 \\
 13 & 0.528 & 0.893 & 2125 \\
 14 & 0.568 & 0.761 & 1875 \\
 15 & 0.611 & 0.638 & 1625 \\
 16 & 0.656 & 0.523 & 1375 \\
 17 & 0.706 & 0.416 & 1125 \\
 18 & 0.76 & 0.315 & 875 \\
 19 & 0.82 & 0.22 & 625 \\
 20 & 0.886 & 0.129 & 375 \\
 21 & 0.96 & 0.042 & 125 \\
 22 & 1 & 0 & 0 \\
\hline
\hline
\end{tabular}
\end{table}

\subsection{Backreaction in the $\Lambda$LTB model} \label{back}

With an LTB perturbation that is exactly matched to a background FLRW metric, one can study in an exact way backreaction, that is, the effect of inhomogeneities on the background dynamics.
Indeed, in this very simplified case, the background expansion is set by construction so that one has to simply look at the mismatch between the background energy densities and the averaged ones.

In order to maximize the effect, one could fill the entire universe with infinitely many spherical patches of different radii and profiles by the Apollonian sphere packing, the three-dimensional extension of the Apollonian gasket, see \citet[Fig.~1]{Marra:2007pm}. For this reason, here we will be concerned with the background dynamics at $r=r_b$ and not at larger radii.

The effect of backreaction can be read from Eq.~\eqref{fred}, which can be rewritten as ($r=r_b$):
\begin{align}
H^2(t) &= {8 \pi G \over 3} \Big ( \langle \rho_m \rangle  +  \rho_\Lambda \Big ) -\left \langle \frac{\cal{R}}{6} \right \rangle +P_{\rm inh} \,, \\
P_{\rm inh} &= {8 \pi G \over 3}  \Big (\rho_m(t) - \langle \rho_m \rangle \Big )
- \frac{k}{a^2}  + \left \langle \frac{\cal{R}}{6} \right \rangle \,,  
\end{align}
where $H(t)$ is the background expansion rate, fixed by construction, and $P_{\rm inh}$ represents `the effects of small-scale inhomogeneities in the universe on the dynamic behavior at the smoothed-out scale' \citep{Ellis1984}.

If there were no backreaction, $P_{\rm inh}=0$, the Friedmann equation would be sourced by the averages of the energy and curvature content of the inhomogeneity.
These are obtained by adopting the actual volume element with curvature, $\d V = 4 \pi Y^2 Y' /\sqrt{1+2 E}$.
However, Einstein's equations are nonlinear so that backreaction gives the correction $P_{\rm inh}$. The correction comes from the fact that, by solving Einstein's equations  {for the LTB metric,} one finds that it is the Euclidean average of the density and curvature that sources the Friedmann equation.
In the case of the density, for example, the correction is proportional to the difference between the invariant mass and $F$---also known as the Misner-Sharp mass \citep[see][for an extensive discussion]{Alfedeel:2009ef}.
As the difference is proportional to the energy function $E \sim \Phi \propto (r_b/ r_{\rm hor})^2$, one concludes that the backreaction of small-scale inhomogeneities into the background expansion is negligible \citep[see, however,][ for a model that does feature large backreaction]{Lavinto:2013exa}.
See  \citet{Sussman:2011na} for a comprehensive discussion of averaging and backreaction in LTB metrics.

\section{$N$-body simulations of a perturbed $\Lambda$LTB model} \label{Nsim}

We now discuss how we simulate the $\Lambda$LTB model.
As said in the Introduction, because of spatial gradients, standard primordial perturbations are coupled at first order so that one may expect a different growth of perturbations even on scales at which the evolution is still linear.
Besides this, simulations are necessary, just as in $\Lambda$CDM, in order to obtain the fully nonlinear structure, which again may be affected by the spatial gradients of an inhomogeneous background.
As we are interested in understanding and modeling these effects, each $\Lambda$LTB simulation will be coupled with the corresponding $\Lambda$CDM one, using the same seed for the initial conditions.
This will allow us to study the differential change in quantities as compared to $\Lambda$CDM, and reduce possible biases caused by the numerical implementation we adopt.

\begin{figure*}
\centering 
\includegraphics[trim={.cm .cm .cm .cm}, clip, width=.8 \textwidth]{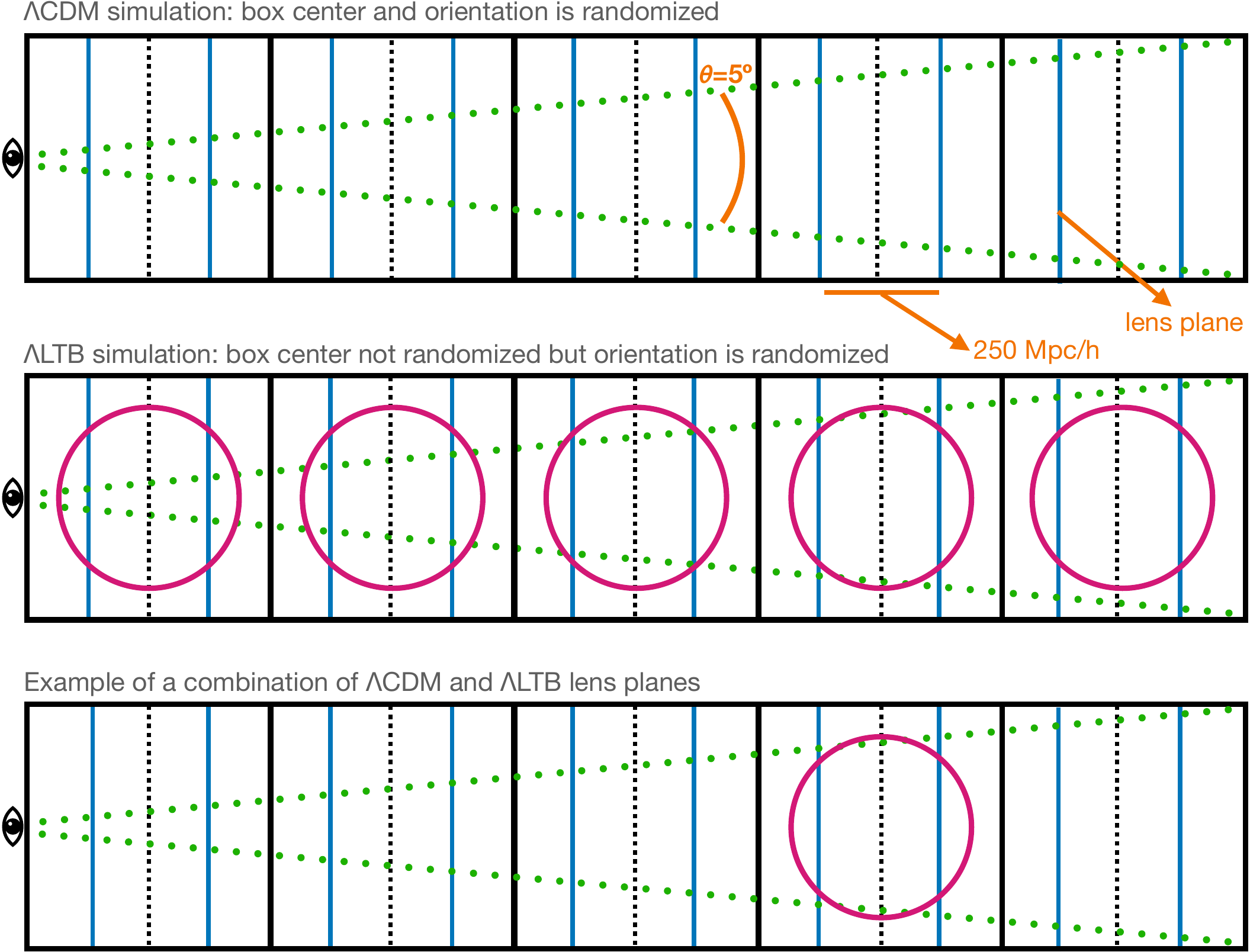}
\caption{Schematic illustration of the lightcone construction, mixing $\Lambda$CDM e $\Lambda$LTB lens planes. The illustration shows the case of the smallest Box 1, see Table~\ref{tab:sims}. See Section~\ref{lensing} for more details.
\label{fig:lens-planes}}
\end{figure*}

\subsection{Early-FLRW initial conditions}

As shown by \citet{Alonso:2010zv,Alonso:2012ds}, one can simulate the $\Lambda$LTB model by feeding standard Newtonian gravity-only $N$-body codes with special early-FLRW initial conditions.
We will give initial conditions at $z_{\rm ini}=49$  so that the LTB perturbation is deep into the linear regime and the spacetime can  be accurately described by a superposition of two kinds of perturbations:
\begin{align}
\delta( t_{\rm ini},\vec x) = \delta^{\rm LTB}(t_{\rm ini},\vec x) + \delta^{\rm gau}( t_{\rm ini},\vec x) \,,
\end{align}
where the first term comes from the spherical LTB perturbation and the second one from the statistically isotropic set of primordial Gaussian perturbations.
As discussed in the Introduction, the LTB initial conditions are non-Gaussian, with phase coupling induced by the presence of the spherical inhomogeneity.

As we have seen, at early times, the potential $\Phi^{\rm LTB}$ induced by the LTB metric  is given by Eq.~\eqref{linpot},  {together with Equations \eqref{energy} and \eqref{profi1}.}
Owing to the Poisson equation in Newtonian gravity, the gravitational potential obeys qualitatively exactly the same differential equation as the displacement potential for the matter field, such that initial positions for particles in a simulation can be set by:
\begin{align}
	\vec x(t_{\rm ini}) = \vec q + \vec\nabla \left( \Phi^{\rm LTB} + \Phi^{\rm gau}  \right) \,.
\end{align}
We will generate  these initial conditions using \texttt{FalconIC} \citep[2017 version,][]{Valkenburg:2015dsa}, a code that extends Lagrangian perturbation theory  to nontrivial theories of gravity.

As the $\Lambda$LTB model does not include radiation, we will neglect radiation in the $N$-body simulation, as well as the effect of neutrinos, including massive neutrinos. In other words, the initial transfer function $T_{\vec k}(z_{\rm ini})$ is obtained by rescaling the one provided by the Boltzmann solver at $z=0$ to the initial redshift via the scale-independent radiationless growth factor $D(z=49)=0.0256745$, where $D(z=0)=1$.
See \citet{Valkenburg:2016xek,Michaux:2020yis} for a thorough discussion on initial conditions for simulations.
\texttt{FalconIC} adopts \texttt{CLASS} \citep{Blas:2011rf}.

As we are adding the LTB perturbation to the standard ones, we need to correctly normalize the sum. Moving to Fourier space, it is:
\begin{align}
	\delta_{\vec k}(t_{\rm ini}) = T_{\vec k}(t_{\rm ini})\left[  \frac{\delta_{\vec k}^{\rm LTB}(t_0)}{ T_{\vec k}(t_{\rm ini})} \frac{\rho_m^{\rm LTB}(r=0, t_{\rm ini})}{\rho_m^{\rm LTB}(r=0, t_{0})} +  \delta_{\vec k}^{\rm gau}(t_{\rm ini}) \right] \,,
\end{align}
where $\delta_{\vec k}^{\rm gau}(t_{\rm ini}) $ is the familiar nearly scale invariant density perturbation as imprinted by inflation, and the fraction of LTB densities guarantees the right normalization of the spherical perturbation.

Finally, as said earlier, each $\Lambda$LTB simulation will be coupled with the corresponding $\Lambda$CDM one, using the same seed for the initial conditions. As the LTB perturbation is added on top of the primordial perturbations, this means that standard large scale structures are preserved and one can factor out cosmic variance when studying the effect of spatial gradients.
In particular, the particles' ID are stable after the addition of the LTB perturbation so that one can even study the effect of the inhomogeneous background at the particle level.

 {Summarizing, the $\Lambda$LTB model is treated as a $\Lambda$CDM model with an extra large-scale perturbation, that is, these are FLRW simulations as far as the $N$-body code is concerned. Specifically, $\apar=\aperp=a$ and $k(r)$ is constant.
Nevertheless, as we will see in Section~\ref{bkgtest}, these simulations exactly show the inhomogeneous background evolution of the $\Lambda$LTB model, on top of which standard perturbations evolve. This allows us to study the effect of spatial gradients on the evolution of perturbations.}

\subsection{Numerical simulation}

Table~\ref{tab:sims} shows the technical details of the simulations, while the cosmology is specified by the parameters of Table~\ref{tab:LLTB-pars}.
The rationale is to explore the parameter space of size and depth of the inhomogeneity. In the table, the first line of each Box section refers to the $\Lambda$CDM simulation with which the $\Lambda$LTB simulations are paired.
For each simulation, 22 snapshots are saved, but only 12 are kept after the generation of the halo catalogs and lens planes. See Table~\ref{tab:snaps} for details.

The simulations only include dark matter, besides the cosmological constant, and were performed using \texttt{OpenGadget3}, a modified version of \texttt{GADGET-2} \citep{Springel:2005mi}.
The number of grid elements of the particle mesh is always twice the number of particles per dimension, and the comoving gravitational softening is chosen according to:
\begin{equation}
\lambda =\left\lbrace \!\!
\begin{array}{lll}
\!\!\!\left( {M_{\rm part} \over 10^9 M_\odot/h} \right)^{1/3} \! 3 (1\!+\!z) \frac{\text{ kpc}}{h}= (1\!+\!z) n_\lambda \, d_{\rm part}   & \text{if} \quad z\le 2   \\
& \\
\!\!\!\left( {M_{\rm part} \over 10^9 M_\odot/h } \right)^{1/3}\! 9  \frac{\text{ kpc}}{h}= 3 \, n_\lambda \, d_{\rm part}  & \text{if}  \quad  z> 2 
\end{array}
\right. \!\!\!,
\label{eq:soft}
\end{equation} 
%
where the comoving interparticle distance $d_{\rm part}$ is given by:
\begin{align}
d_{\rm part}&=\frac{L_{\rm box}}{N_{\rm part}^{1/3}}  \,,
\end{align}
and the constant $n_\lambda$
by: 
\begin{align}
n_\lambda &= \left ( \frac{\rho_{c0} h^{-2}\,\Omega_{m0}}{10^9 M_\odot/h} \right) ^{1/3} 3 \text{ kpc}/h \simeq \frac{1}{76.3} \,.
\end{align}

The rms initial displacement, generated at $z=49$, is given by:
\begin{align}
\sqrt{\left \langle \left (x-x_{\rm grid} \right)^2 \right \rangle} = n_{\rm disp} \, d_p \,,
\end{align}
%
where $x$ is the coordinate of the particle and $x_{\rm grid}$ is the coordinate of the grid. As it is $n_{\rm disp}  \sim 0.2 \ll 1$,  initial conditions were given early enough so that there is no risk of shell crossing.%
\footnote{Note that, because of bulk motion, this is a conservative check.}

As shown by the last column of Table~\ref{tab:sims}, even the most nonlinear $\Lambda$LTB simulation took just $\approx$50\% more CPU hours as compared to the corresponding $\Lambda$CDM simulation.
 {The increase in CPU time is due to the fact that the LTB inhomogeneity distorts the large-scale structure, increasing the nonlinearity in some regions, while decreasing it in others. The more nonlinear regions require more integration steps, and therefore a longer computational time.}
In Appendix~\ref{ap:perfo} we show how the execution of a simulation with \texttt{OpenGadget3} is affected by the background inhomogeneity, again highlighting a minor impact,  {except for the most nonlinear cases.}
Concluding, one can achieve the same resolution of standard $\Lambda$CDM simulations in approximately the same CPU time.

\subsection{Halo catalog}

We obtained the halo catalogs and merger trees with \texttt{Rockstar} \citep[v0.99.9-RC3,][]{Behroozi:2011ju}, for the 21 snapshots from $z=0$ to $z=4.2$ that are described in Table~\ref{tab:snaps}.
In order to have a comprehensive characterization, for each halo, 40 physical properties are saved.
In particular, the masses were computed using strict spherical overdensity masses according to the definitions $M_{\rm vir}$, $M_{200m}$, $M_{200c}$, $M_{500c}$ and $M_{2500c}$.
Finally, although we set the minimum halo size of 20 particles, we only consider halos with 50 or more particles as suggested by the results of \citet[]{Leroy:2020fzc}.

\begin{figure*}
\centering 
\includegraphics[trim={.cm .cm .cm .cm}, clip, width= .9 \textwidth]{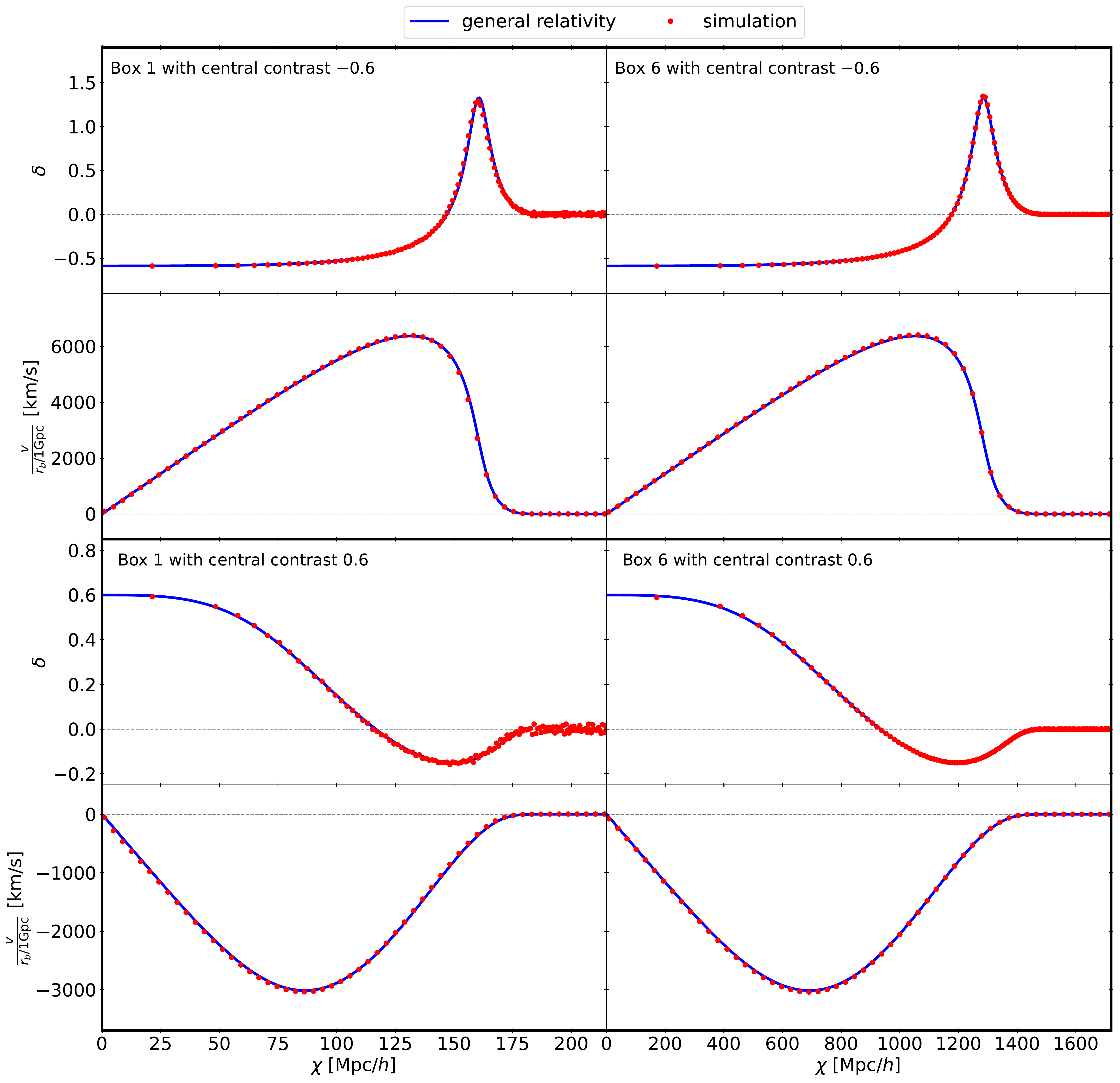}
\caption{Density and velocity profiles at $z=0$ for the smallest and largest boxes with the largest contrasts, see Table~\ref{tab:sims}.
In order to precisely test the background dynamics, we simulate an inhomogeneous universe without the standard primordial Gaussian perturbations ($A_s \approx 0$).
The inhomogeneous (Newtonian) $N$-body simulations perfectly follow the general relativistic solution. Furthermore, thanks to the scale invariance, their re-scaled evolution is the same.
See Section~\ref{bkgtest} for more details.
\label{fig:bkg-test-nopert}}
\end{figure*}

\subsection{Gravitational lensing} \label{lensing}

We obtained the lens planes and maps with \texttt{SLICER}.%
\footnote{\href{https://github.com/TiagoBsCastro/SLICER}{github.com/TiagoBsCastro/SLICER} (2021 version)}
Starting from the observer, the lens planes are computed every $250 \text{Mpc}/h$. In order to minimize the extrapolation on the particle positions and to probe the interior part of the smaller Box 1 simulations, the snapshots are saved at the redshifts that correspond to $d_{C,i}=( i \times 250 -125) \text{ Mpc}/h$, as summarized in Table~\ref{tab:snaps}.

We generated 21 lens planes of 2048$^2$ pixels with a $5^\circ \times 5^\circ$ field of view up to $z=4.2$, with a resolution of 8.8 arcsec.
For the $\Lambda$CDM simulations the orientations and the centers of the boxes were randomized. A total of 10 lightcones were produced in order to reduce sample variance. For the $\Lambda$LTB simulations, the centers were not randomized in order to preserve the LTB symmetry. The orientations were randomized in order to obtain, also in this case, 10 lightcones. As the background evolution is the same by construction, one can then place one or more LTB inhomogeneities at various redshifts, padding the remaining lightcones with the lens planes from the $\Lambda$CDM simulation. In other words, one may or may not have periodicity in the line-of-sight distribution of inhomogeneities.
Figure~\ref{fig:lens-planes} illustrates the lightcone construction.
For any such combination one can then use \texttt{SLICER} to obtain convergence, shear and lensing potential maps.

\section{Results} \label{results}

\begin{figure*}
\centering 
\includegraphics[trim={.cm .cm .cm .cm}, clip, width= \textwidth]{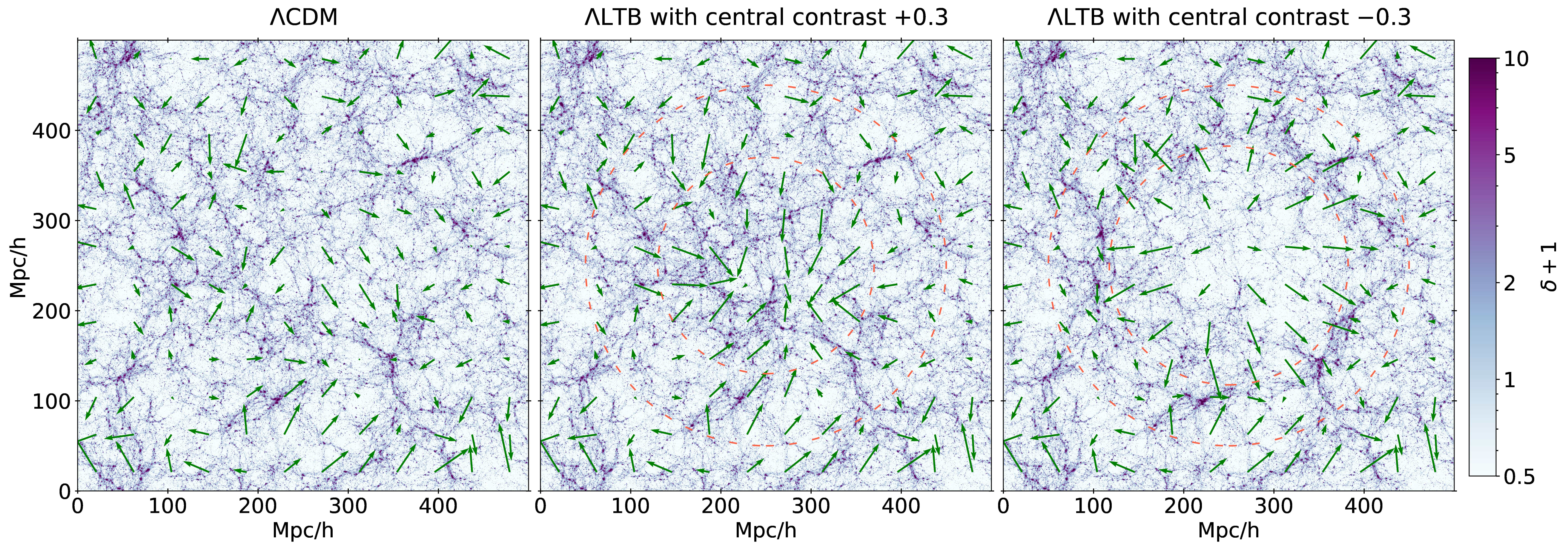}
\includegraphics[trim={.cm .cm .cm .cm}, clip, width=.93 \textwidth]{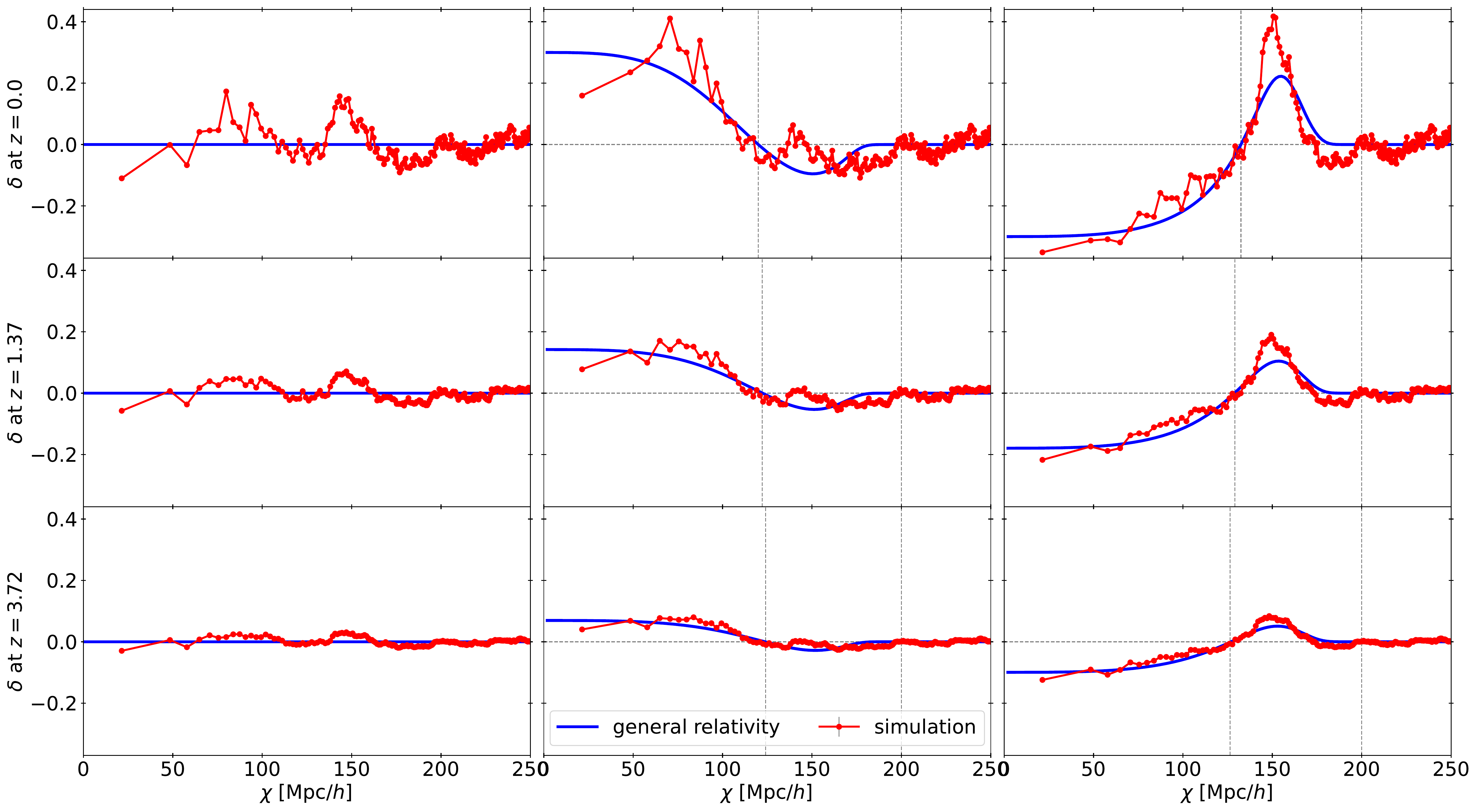}
\hspace{1cm}
\caption{
The first row shows the large-scale structure of Box 1 at $z=0$ of the overdense (middle panel) and underdense (right panel) $\Lambda$LTB models together with the corresponding $\Lambda$CDM model (left panel).
The larger thin dashed circle marks the boundary $r_b$ of the $\Lambda$LTB inhomogeneity and the smaller one the radius $r_t$ at which $\delta(t_0, r_t)=0$, which marks the transition from the central under- or overdensity to the compensating  over- or underdense shell.
The arrows show the velocity field. The density and velocity fields are  obtained from the projection of the slice through the center, whose thickness is a fifth of the box side. 
One can see how the large-scale structure is identical outside the inhomogeneity, but it is distorted by the inhomogeneous bulk flow inside the LTB structure.
The last three rows show the evolution of the radial profile, from $z=3.7$ to $z=0$ (Poissonian errors are negligible). Also shown is the general relativistic solution  {given by the LTB solution of Section~\ref{LTBmodel}.} The vertical lines mark $r_b$ and the smaller $r_t$. While $r_b$ is fixed in comoving coordinates, $r_t$ moves because of the peculiar velocity of the LTB structure.
\label{fig:snaps1}}
\end{figure*}

\begin{figure*}
\centering 
\includegraphics[trim={.cm .cm .cm .cm}, clip, width= \textwidth]{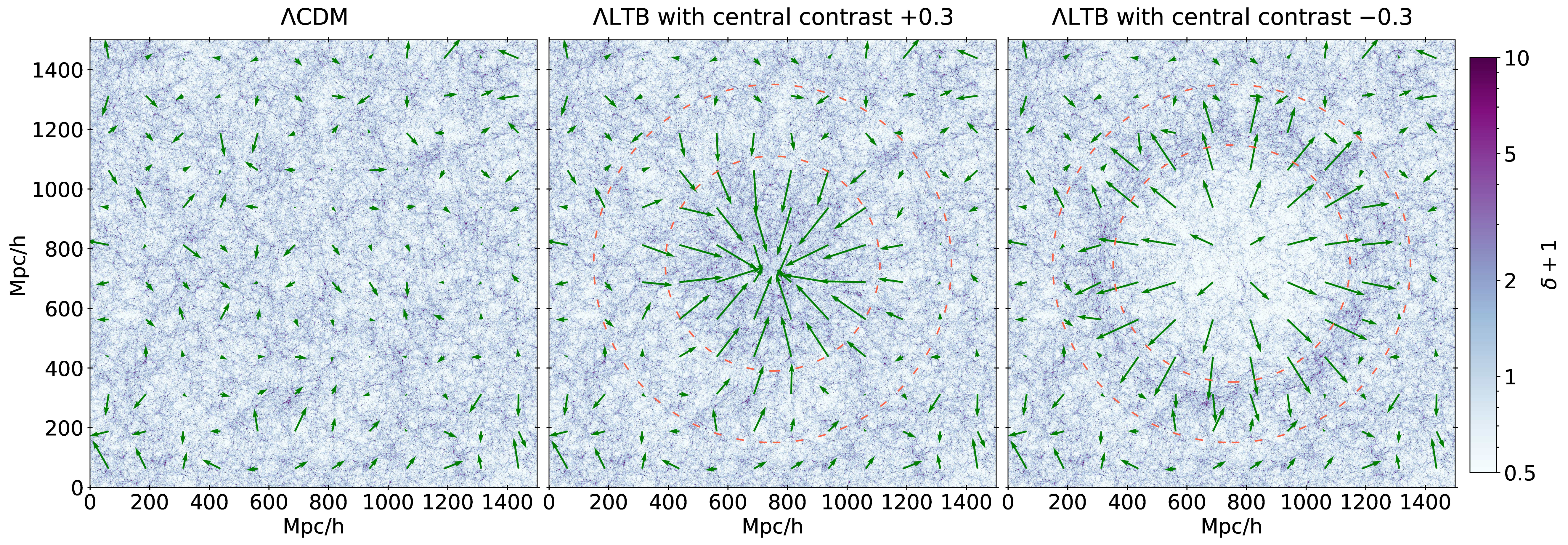}
\includegraphics[trim={.cm .cm .cm .cm}, clip, width=.93 \textwidth]{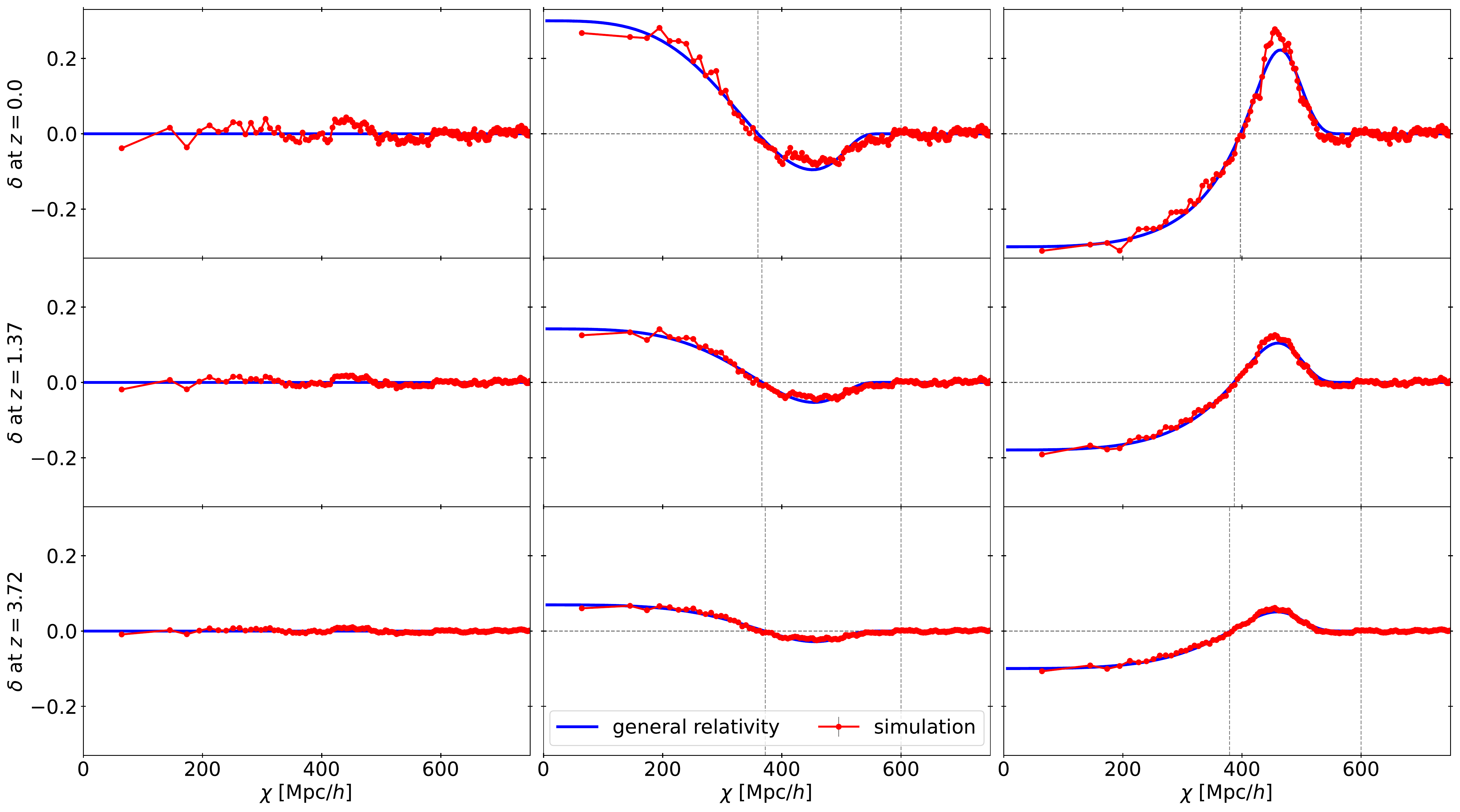}
\hspace{1cm}
\caption{As Figure~\ref{fig:snaps1} but for Box 3. Because of the larger size, small-scale perturbations average out and the structure is more visible.
\label{fig:snaps3}}
\end{figure*}

We now show the results of the simulations.
 {Table~\ref{tab:sims} lists the LTB parameters that we adopt, together with a few summary statistics such as the total number of halos and the masses of the most massive halos.
The cosmology is specified in Table~\ref{tab:LLTB-pars}.
We adopt a grid in the LTB parameter space that covers inhomogeneities with radius $r_b$ from 200 Mpc/$h$ to 1.6 Gpc/$h$ and contrasts $\delta_0$ from $\pm 0.1$ to $\pm 0.6$, for a total of 68 models.
The rationale behind this choice is to systematically explore the phenomenology of $\Lambda$LTB models in order to accurately understand how the growth of perturbations changes as a function of the LTB parameters.
}

 {
It is worth mentioning that present-day observations rule out part of the models of Table~\ref{tab:sims}.
As anticipated in the introduction, structures at the last scattering surface or along the line of sight (see the lightcone construction of Figure~\ref{fig:lens-planes}) cannot have nonlinear contrasts.
Considering that we use the contrast at the center $\delta_0$ and that the average contrast $\Delta(t_0, r_t)$ is lower than $\delta_0$ (see Figure~\ref{example-t0}, second panel from the top), the models with $|\delta_0| > 0.3 $ are ruled out.
On the other hand, if the observer is at the center, the models with $|\delta_0| \ge 0.3 $ and $r_b \ge 400$ Mpc/$h$ are excluded as discussed in Appendix~\ref{ap:constraints}, where the latest constraints by \citet[]{Camarena:2021mjr} are shown.
Despite these observational constraints, we nevertheless simulate the models with $|\delta_0| \ge 0.3 $ in order to test different approximation schemes for the growth of perturbations and their limits, which is the subject of a forthcoming work.
In other words, we simulate a larger than observationally allowed parameter space in order to obtain a better modeling of linear and nonlinear structures.
In particular, the main effects caused by spatial gradients will be clearly evident in the most nonlinear models, allowing us to disentangle the signal from the noise with just one $N$-body realization (one  seed). 
We anyway sample more finely the observationally relevant parameter space with $|\delta_0| \le 0.3 $.
}

Before presenting the evolution of the $\Lambda$LTB large-scale structure we will discuss the validation of the $\Lambda$LTB background evolution.
The validation of the $\Lambda$CDM simulations is discussed in Appendix~\ref{ap:LCDMtest}, where we show that this first set of BEHOMO simulations have power spectrum and halo mass function accurate at the 5\% level.
As said before, the $\Lambda$LTB and $\Lambda$CDM simulations are paired so that numerical errors should approximately factor out when considering suitable ratios of relevant quantities.

\subsection{Validation of the $\Lambda$LTB background dynamics} \label{bkgtest}

We validate the large-scale dynamics produced by the inhomogeneous (Newtonian) $N$-body simulations via the exact $\Lambda$LTB solution of General Relativity described in Section~\ref{LTBmodel}. As we are not interested in small-scale dynamics, here we will adopt lower-resolution simulations.
Specifically, we consider the smallest Box 1 with $256^3$ particles and the largest Box 6 with $1024^3$ particles, both with the largest contrasts of $\delta_0 = \pm 0.6$.
Box 1 has a side of 500 Mpc/$h$ and inhomogeneity radius of $r_b=200$ Mpc/$h$, while Box 6 has side of 4000 Mpc/$h$ and $r_b=1600$ Mpc/$h$, see Table~\ref{tab:sims}.
In order to precisely test the background dynamics, we simulate an inhomogeneous universe without the standard primordial Gaussian perturbations, that is, we adopt a very small amplitude of the primordial power spectrum, $A_s \approx 0$.

Fig.~\ref{fig:bkg-test-nopert} compares the density and  velocity profiles at $z=0$.
Using Eq.~\eqref{velo}, we  connect the longitudinal velocities $v_{\parallel}$ from the simulation to the perpendicular Hubble rate $\Hperp$:
\begin{align}
v_{\parallel}(t,\chi) = a(t) \, \chi \, \big[ \Hperp(t,r) - H(t)   \big]  \,.
\end{align}
%
One can see that the $N$-body simulation produces a background evolution that perfectly follows the general relativistic one, even for inhomogeneities whose size is comparable with the Hubble radius.
This is ultimately due to the fact that the $\Lambda$LTB dynamics is scale invariant thanks to the symmetries of the LTB metric, see Section~\ref{scale-inv}.

As the LTB inhomogeneity is comparable in size with the box ($r_b=0.4 \, L$), one could wonder if such a large inhomogeneity would self-interact via the periodic boundary conditions.
The answer is no because the LTB metric is matched at a finite radius rather than asymptotically \citep[see][for the latter case]{Alonso:2010zv}. Therefore, a particle outside the inhomogeneity, such as the LTB structure itself with respect to its mirror LTB image, would locally experience the FLRW background and, thanks to the shell theorem (valid because of spherical symmetry), the same gravitational force as if there were no mirror images. Figure~\ref{fig:bkg-test-nopert} confirms this.

The profile that we adopted is smooth at the center so that no artifacts are introduced when using the particle mesh part of  \texttt{OpenGadget3}, as instead observed by \citet{Alonso:2010zv} when adopting a cuspy profile.
Using particle mesh is especially convenient (computationally less expensive)  at early times when the tree algorithm struggles to deliver an accurate computation of the gravitational force~\citep[][]{Springel:2020plp}.

\subsection{The large-scale structure in an inhomogeneous universe} \label{snaps}

In Figure~\ref{fig:snaps1} and~\ref{fig:snaps3} we show, for Box 1 and Box 3, the large-scale structure of the $\Lambda$LTB model as compared with the corresponding one of the $\Lambda$CDM model.
While the larger circle marks the boundary $r_b$ of the $\Lambda$LTB inhomogeneity, the smaller one marks the transition from the central under- or overdensity to the compensating surrounding over- or underdense shell. 
Also shown is the velocity field.

Near the center one can see how  matter structures are amplified or reduced, when a central overdensity or underdensity is present, respectively.
It is evident that the inhomogeneity causes a deformation of the positions and velocities of the corresponding particles in the $\Lambda$CDM simulation, and that such deformation disappears outside the inhomogeneity.
In other words, the effect of the inhomogeneous background is to deform the large-scale structure of the $\Lambda$CDM model. Indeed, as pointed out earlier, the LTB perturbation is added on top of the primordial perturbations so that the structure of the cosmic web is preserved.
Statistically, this will allow us to factor out cosmic variance when studying the effect of spatial gradients on observables.

The bottom panels of Figs.~\ref{fig:snaps1} and~\ref{fig:snaps3} show the evolution of the radial profiles at $z=0$, 1.37 and 3.72 (from upper to lower panels). Also shown with the blue curves is the general relativistic solution  {given by the LTB solution presented in Section~\ref{LTBmodel}.} The vertical lines mark the inhomogeneity radius $r_b$ and the smaller transition radius $r_t$. While $r_b$ is fixed in comoving coordinates, $r_t$ moves because of the peculiar velocity of the LTB structure.
The LTB evolution of Box 1 and Box 3 is identical except for the scales involved, but the inhomogeneity is more clearly seen in Box 3 because, thanks to its larger size, small-scale perturbations average out.

\section{Conclusions} \label{conclusions}

In this paper we presented the BEHOMO project (cosmology {\bf be}yond {\bf homo}geneity and isotropy) and its first suite of simulations.
The goal is to study, via the methods of numerical cosmology, the Universe without assuming large-scale homogeneity and isotropy.
In order to present a viable program, we consider early-FLRW cosmologies, which are models that,  at early times, are near-FLRW  so that the standard inflationary paradigm is maintained and the physics that leads to the CMB remains unchanged. 
As a first realization of these inhomogeneous cosmologies, we adopted the $\Lambda$LTB spherical model, on which the simulations here presented are based.

After a comprehensive review of the $\Lambda$LTB model, we described the numerical implementation of the $\Lambda$LTB simulations.
The data products consist of 11 snapshots between redshifts $z=0$ and $z=3.7$ for each of the 68 simulations that have been performed, together with halo catalogs and lens planes relative to 21 snapshots, between redshift 0 and 4.2, for a total of approximately 180 TB of data.
This is the first set of simulations of the $\Lambda$LTB model ever produced. In particular, we chose the inhomogeneity profile so that these simulations do not suffer from any spurious artifacts.
Indeed, these Newtonian $N$-body simulations can perfectly reproduce the general relativistic evolution even for deep Hubble-sized inhomogeneities.

With these data products we plan to study in forthcoming papers the growth of perturbations at the linear and nonlinear level, gravitational lensing, cluster abundances and proprieties, and many other applications which we invite the scientific community to propose. Data can be obtained upon request.

After the exploitation of this first suite of simulations, the BEHOMO project will consider more realistic scenarios.
One may consider exact solutions such as the quasi-spherical Szekeres model \citep{1975CMaPh..41...55S}, which features a dipole inhomogeneity instead of a spherical one \citep{Bolejko:2006my}, or a more general inhomogeneous model to be solved via numerical GR codes.
The ultimate goal is to constrain inhomogeneous models with present and future background and perturbation observables.
Further information is available at \href{https://valerio-marra.github.io/BEHOMO-project}{valerio-marra.github.io/BEHOMO-project}.


\begin{acknowledgements}
Warm thanks to Wessel Valkenburg for help during the early stages of this project and for sharing \texttt{VoidDistances2020} and \texttt{FalconIC}.
It is a pleasure to thank Klaus Dolag for sharing \texttt{OpenGadget3}, and Alessio Notari, Miguel Quartin and Emiliano Sefusatti for useful discussions.\\
VM thanks CNPq (Brazil) and FAPES (Brazil) for partial financial support.
This project has received funding from the European Union’s Horizon 2020 research and innovation programme under the Marie Skłodowska-Curie grant agreement No \href{https://cordis.europa.eu/project/id/888258}{888258}.
TC is supported by the fare Miur grant `ClustersXEuclid' R165SBKTMA.
TC and SB are supported by the INFN IN-DARK PD51 grant.
DC thanks CAPES for financial support.
AR acknowledges support from the grant PRIN-MIUR 2017 WSCC32.\\
We acknowledge the use of the HOTCAT computing infrastructure of the Astronomical Observatory of Trieste of the National Institute for Astrophysics (INAF, Italy) \citep[see][]{2020ASPC..527..303B,2020ASPC..527..307T}.
We acknowledge the computing centre  of Cineca and INAF, under the coordination of the ``Accordo Quadro MoU per lo svolgimento di attività congiunta di ricerca Nuove frontiere in Astrofisica: HPC e Data Exploration di nuova generazione,'' for the availability of computing resources and support.
We  acknowledge the use of the Santos Dumont supercomputer of the National Laboratory of Scientific Computing (LNCC, Brazil). 
\end{acknowledgements}

\bibliographystyle{aaArxivDoi}
\bibliography{biblio}

\begin{thebibliography}{109}
\expandafter\ifx\csname natexlab\endcsname\relax\def\natexlab#1{#1}\fi

\bibitem[{Abate {et~al.}(2012)}]{LSSTDarkEnergyScience:2012kar}
Abate, A. {et~al.} , [\href{https://arxiv.org/abs/1211.0310}{1211.0310}].

\bibitem[{Abbott {et~al.}(2022)}]{DES:2021wwk}
Abbott, T. M.~C. {et~al.} 2022,
  \href{https://doi.org/10.1103/PhysRevD.105.023520}{Phys. Rev. D}, 105,
  023520, [\href{https://arxiv.org/abs/2105.13549}{2105.13549}].

\bibitem[{Abdalla {et~al.}(2022)}]{Abdalla:2022yfr}
Abdalla, E. {et~al.} 2022,
  \href{https://doi.org/10.1016/j.jheap.2022.04.002}{JHEAp}, 34, 49,
  [\href{https://arxiv.org/abs/2203.06142}{2203.06142}].

\bibitem[{Adamek {et~al.}(2020)Adamek, Barrera-Hinojosa, Bruni, Li, Macpherson,
  \& Mertens}]{Adamek:2020jmr}
Adamek, J., Barrera-Hinojosa, C., Bruni, M., {et~al.} 2020,
  \href{https://doi.org/10.1088/1361-6382/ab939b}{Class. Quant. Grav.}, 37,
  154001, [\href{https://arxiv.org/abs/2003.08014}{2003.08014}].

\bibitem[{Adamek {et~al.}(2019)Adamek, Clarkson, Daverio, Durrer, \&
  Kunz}]{Adamek:2017mzb}
Adamek, J., Clarkson, C., Daverio, D., Durrer, R., \& Kunz, M. 2019,
  \href{https://doi.org/10.1088/1361-6382/aaeca5}{Class. Quant. Grav.}, 36,
  014001, [\href{https://arxiv.org/abs/1706.09309}{1706.09309}].

\bibitem[{Adamek {et~al.}(2016)Adamek, Daverio, Durrer, \&
  Kunz}]{Adamek:2016zes}
Adamek, J., Daverio, D., Durrer, R., \& Kunz, M. 2016,
  \href{https://doi.org/10.1088/1475-7516/2016/07/053}{JCAP}, 07, 053,
  [\href{https://arxiv.org/abs/1604.06065}{1604.06065}].

\bibitem[{Aghamousa {et~al.}(2016)}]{DESI:2016fyo}
Aghamousa, A. {et~al.} , [\href{https://arxiv.org/abs/1611.00036}{1611.00036}].

\bibitem[{Aghanim {et~al.}(2020)}]{Planck:2018vyg}
Aghanim, N. {et~al.} 2020,
  \href{https://doi.org/10.1051/0004-6361/201833910}{Astron. Astrophys.}, 641,
  A6, [Erratum: Astron.Astrophys. 652, C4 (2021)],
  [\href{https://arxiv.org/abs/1807.06209}{1807.06209}].

\bibitem[{Alfedeel \& Hellaby(2010)}]{Alfedeel:2009ef}
Alfedeel, A. A.~H. \& Hellaby, C. 2010,
  \href{https://doi.org/10.1007/s10714-010-0971-y}{Gen. Rel. Grav.}, 42, 1935,
  [\href{https://arxiv.org/abs/0906.2343}{0906.2343}].

\bibitem[{Alonso {et~al.}(2012)Alonso, Garcia-Bellido, Haugboelle, \&
  Knebe}]{Alonso:2012ds}
Alonso, D., Garcia-Bellido, J., Haugboelle, T., \& Knebe, A. 2012,
  \href{https://doi.org/10.1016/j.dark.2012.08.001}{Phys. Dark Univ.}, 1, 24,
  [\href{https://arxiv.org/abs/1204.3532}{1204.3532}].

\bibitem[{Alonso {et~al.}(2010)Alonso, Garcia-Bellido, Haugbolle, \&
  Vicente}]{Alonso:2010zv}
Alonso, D., Garcia-Bellido, J., Haugbolle, T., \& Vicente, J. 2010,
  \href{https://doi.org/10.1103/PhysRevD.82.123530}{Phys. Rev. D}, 82, 123530,
  [\href{https://arxiv.org/abs/1010.3453}{1010.3453}].

\bibitem[{Amendola {et~al.}(2018)}]{Amendola:2016saw}
Amendola, L. {et~al.} 2018,
  \href{https://doi.org/10.1007/s41114-017-0010-3}{Living Rev. Rel.}, 21, 2,
  [\href{https://arxiv.org/abs/1606.00180}{1606.00180}].

\bibitem[{Angulo {et~al.}(2021)Angulo, Zennaro, Contreras, Aric\`o,
  Pellejero-Iba\~nez, \& St\"ucker}]{Angulo:2020vky}
Angulo, R.~E., Zennaro, M., Contreras, S., {et~al.} 2021,
  \href{https://doi.org/10.1093/mnras/stab2018}{Mon. Not. Roy. Astron. Soc.},
  507, 5869, [\href{https://arxiv.org/abs/2004.06245}{2004.06245}].

\bibitem[{Barrera-Hinojosa \&
  Li(2020{\natexlab{a}})}]{Barrera-Hinojosa:2019mzo}
Barrera-Hinojosa, C. \& Li, B. 2020{\natexlab{a}},
  \href{https://doi.org/10.1088/1475-7516/2020/01/007}{JCAP}, 01, 007,
  [\href{https://arxiv.org/abs/1905.08890}{1905.08890}].

\bibitem[{Barrera-Hinojosa \&
  Li(2020{\natexlab{b}})}]{Barrera-Hinojosa:2020arz}
Barrera-Hinojosa, C. \& Li, B. 2020{\natexlab{b}},
  \href{https://doi.org/10.1088/1475-7516/2020/04/056}{JCAP}, 04, 056,
  [\href{https://arxiv.org/abs/2001.07968}{2001.07968}].

\bibitem[{Behroozi {et~al.}(2013)Behroozi, Wechsler, \& Wu}]{Behroozi:2011ju}
Behroozi, P.~S., Wechsler, R.~H., \& Wu, H.-Y. 2013,
  \href{https://doi.org/10.1088/0004-637X/762/2/109}{Astrophys. J.}, 762, 109,
  [\href{https://arxiv.org/abs/1110.4372}{1110.4372}].

\bibitem[{Bentivegna \& Bruni(2016)}]{Bentivegna:2015flc}
Bentivegna, E. \& Bruni, M. 2016,
  \href{https://doi.org/10.1103/PhysRevLett.116.251302}{Phys. Rev. Lett.}, 116,
  251302, [\href{https://arxiv.org/abs/1511.05124}{1511.05124}].

\bibitem[{{Bertocco} {et~al.}(2020){Bertocco}, {Goz}, {Tornatore}, {Ragagnin},
  {Maggio}, {Gasparo}, {Vuerli}, {Taffoni}, \&
  {Molinaro}}]{2020ASPC..527..303B}
{Bertocco}, S., {Goz}, D., {Tornatore}, L., {et~al.} 2020, in Astronomical
  Society of the Pacific Conference Series, ed. R.~{Pizzo}, E.~R. {Deul}, J.~D.
  {Mol}, J.~{de Plaa}, \& H.~{Verkouter}, Vol. 527, 303

\bibitem[{Biswas {et~al.}(2007)Biswas, Mansouri, \& Notari}]{Biswas:2006ub}
Biswas, T., Mansouri, R., \& Notari, A. 2007,
  \href{https://doi.org/10.1088/1475-7516/2007/12/017}{JCAP}, 12, 017,
  [\href{https://arxiv.org/abs/astro-ph/0606703}{astro-ph/0606703}].

\bibitem[{Biswas \& Notari(2008)}]{Biswas:2007gi}
Biswas, T. \& Notari, A. 2008,
  \href{https://doi.org/10.1088/1475-7516/2008/06/021}{JCAP}, 06, 021,
  [\href{https://arxiv.org/abs/astro-ph/0702555}{astro-ph/0702555}].

\bibitem[{Biswas {et~al.}(2010)Biswas, Notari, \& Valkenburg}]{Biswas:2010xm}
Biswas, T., Notari, A., \& Valkenburg, W. 2010,
  \href{https://doi.org/10.1088/1475-7516/2010/11/030}{JCAP}, 11, 030,
  [\href{https://arxiv.org/abs/1007.3065}{1007.3065}].

\bibitem[{Blas {et~al.}(2011)Blas, Lesgourgues, \& Tram}]{Blas:2011rf}
Blas, D., Lesgourgues, J., \& Tram, T. 2011,
  \href{https://doi.org/10.1088/1475-7516/2011/07/034}{JCAP}, 1107, 034,
  [\href{https://arxiv.org/abs/1104.2933}{1104.2933}].

\bibitem[{Bolejko(2007)}]{Bolejko:2006my}
Bolejko, K. 2007, \href{https://doi.org/10.1103/PhysRevD.75.043508}{Phys. Rev.
  D}, 75, 043508,
  [\href{https://arxiv.org/abs/astro-ph/0610292}{astro-ph/0610292}].

\bibitem[{Bolejko {et~al.}(2011)Bolejko, Celerier, \&
  Krasinski}]{Bolejko:2011jc}
Bolejko, K., Celerier, M.-N., \& Krasinski, A. 2011,
  \href{https://doi.org/10.1088/0264-9381/28/16/164002}{Class. Quant. Grav.},
  28, 164002, [\href{https://arxiv.org/abs/1102.1449}{1102.1449}].

\bibitem[{Bondi(1947)}]{Bondi:1947fta}
Bondi, H. 1947, \href{https://doi.org/10.1093/mnras/107.5-6.410}{Mon. Not. Roy.
  Astron. Soc.}, 107, 410.

\bibitem[{Bonoli {et~al.}(2021)}]{Bonoli:2020ciz}
Bonoli, S. {et~al.} 2021,
  \href{https://doi.org/10.1051/0004-6361/202038841}{Astron. Astrophys.}, 653,
  A31, [\href{https://arxiv.org/abs/2007.01910}{2007.01910}].

\bibitem[{Bonvin {et~al.}(2006)Bonvin, Durrer, \& Gasparini}]{Bonvin:2005ps}
Bonvin, C., Durrer, R., \& Gasparini, M.~A. 2006,
  \href{https://doi.org/10.1103/PhysRevD.85.029901}{Phys. Rev. D}, 73, 023523,
  [Erratum: Phys.Rev.D 85, 029901 (2012)],
  [\href{https://arxiv.org/abs/astro-ph/0511183}{astro-ph/0511183}].

\bibitem[{Braun {et~al.}(2015)Braun, Bourke, Green, Keane, \&
  Wagg}]{Braun:2015zta}
Braun, R., Bourke, T., Green, J.~A., Keane, E., \& Wagg, J. 2015,
  \href{https://doi.org/10.22323/1.215.0174}{PoS}, AASKA14, 174.

\bibitem[{{Bruni} {et~al.}(1995){Bruni}, {Matarrese}, \&
  {Pantano}}]{1995ApJ...445..958B}
{Bruni}, M., {Matarrese}, S., \& {Pantano}, O. 1995,
  \href{https://doi.org/10.1086/175755}{\apj}, 445, 958,
  [\href{https://arxiv.org/abs/astro-ph/9406068}{astro-ph/9406068}].

\bibitem[{Buchert {et~al.}(2015)}]{Buchert:2015iva}
Buchert, T. {et~al.} 2015,
  \href{https://doi.org/10.1088/0264-9381/32/21/215021}{Class. Quant. Grav.},
  32, 215021, [\href{https://arxiv.org/abs/1505.07800}{1505.07800}].

\bibitem[{Bull {et~al.}(2012)Bull, Clifton, \& Ferreira}]{Bull:2011wi}
Bull, P., Clifton, T., \& Ferreira, P.~G. 2012,
  \href{https://doi.org/10.1103/PhysRevD.85.024002}{Phys. Rev. D}, 85, 024002,
  [\href{https://arxiv.org/abs/1108.2222}{1108.2222}].

\bibitem[{Cai {et~al.}(2010)Cai, Cole, Jenkins, \& Frenk}]{Cai:2010hx}
Cai, Y.-C., Cole, S., Jenkins, A., \& Frenk, C.~S. 2010,
  \href{https://doi.org/10.1111/j.1365-2966.2010.16946.x}{Mon. Not. Roy.
  Astron. Soc.}, 407, 201, [\href{https://arxiv.org/abs/1003.0974}{1003.0974}].

\bibitem[{Camarena \& Marra(2018)}]{Camarena:2018nbr}
Camarena, D. \& Marra, V. 2018,
  \href{https://doi.org/10.1103/PhysRevD.98.023537}{Phys. Rev.}, D98, 023537,
  [\href{https://arxiv.org/abs/1805.09900}{1805.09900}].

\bibitem[{Camarena {et~al.}(2021)Camarena, Marra, Sakr, \&
  Clarkson}]{Camarena:2021mjr}
Camarena, D., Marra, V., Sakr, Z., \& Clarkson, C. 2021,
  \href{https://doi.org/10.1093/mnras/stab3077}{Mon. Not. Roy. Astron. Soc.},
  509, 1291, [\href{https://arxiv.org/abs/2107.02296}{2107.02296}].

\bibitem[{Camarena {et~al.}(2022)Camarena, Marra, Sakr, \&
  Clarkson}]{Camarena:2022iae}
Camarena, D., Marra, V., Sakr, Z., \& Clarkson, C. ,
  [\href{https://arxiv.org/abs/2205.05422}{2205.05422}].

\bibitem[{Castro {et~al.}(2022)}]{Castrone}
Castro, T. {et~al.} 2022, In preparation

\bibitem[{Chung \& Romano(2006)}]{Chung:2006xh}
Chung, D. J.~H. \& Romano, A.~E. 2006,
  \href{https://doi.org/10.1103/PhysRevD.74.103507}{Phys. Rev. D}, 74, 103507,
  [\href{https://arxiv.org/abs/astro-ph/0608403}{astro-ph/0608403}].

\bibitem[{Clarkson(2012)}]{Clarkson:2012bg}
Clarkson, C. 2012, \href{https://doi.org/10.1016/j.crhy.2012.04.005}{Comptes
  Rendus Physique}, 13, 682,
  [\href{https://arxiv.org/abs/1204.5505}{1204.5505}].

\bibitem[{Clarkson {et~al.}(2009)Clarkson, Clifton, \&
  February}]{Clarkson:2009sc}
Clarkson, C., Clifton, T., \& February, S. 2009,
  \href{https://doi.org/10.1088/1475-7516/2009/06/025}{JCAP}, 06, 025,
  [\href{https://arxiv.org/abs/0903.5040}{0903.5040}].

\bibitem[{Clarkson {et~al.}(2011)Clarkson, Ellis, Larena, \&
  Umeh}]{Clarkson:2011zq}
Clarkson, C., Ellis, G., Larena, J., \& Umeh, O. 2011,
  \href{https://doi.org/10.1088/0034-4885/74/11/112901}{Rept. Prog. Phys.}, 74,
  112901, [\href{https://arxiv.org/abs/1109.2314}{1109.2314}].

\bibitem[{Clarkson \& Regis(2011)}]{Clarkson:2010ej}
Clarkson, C. \& Regis, M. 2011,
  \href{https://doi.org/10.1088/1475-7516/2011/02/013}{JCAP}, 02, 013,
  [\href{https://arxiv.org/abs/1007.3443}{1007.3443}].

\bibitem[{Coles \& Lucchin(2002)}]{coles2003cosmology}
Coles, P. \& Lucchin, P. 2002, Cosmology: The Origin and Evolution of Cosmic
  Structure (Wiley)

\bibitem[{Dakin {et~al.}(2021)Dakin, Hannestad, \& Tram}]{Dakin:2021ivb}
Dakin, J., Hannestad, S., \& Tram, T. ,
  [\href{https://arxiv.org/abs/2112.01508}{2112.01508}].

\bibitem[{Daverio {et~al.}(2019)Daverio, Dirian, \& Mitsou}]{Daverio:2019gql}
Daverio, D., Dirian, Y., \& Mitsou, E. 2019,
  \href{https://doi.org/10.1088/1475-7516/2019/10/065}{JCAP}, 10, 065,
  [\href{https://arxiv.org/abs/1904.07841}{1904.07841}].

\bibitem[{Dunsby {et~al.}(2010)Dunsby, Goheer, Osano, \& Uzan}]{Dunsby:2010ts}
Dunsby, P., Goheer, N., Osano, B., \& Uzan, J.-P. 2010,
  \href{https://doi.org/10.1088/1475-7516/2010/06/017}{JCAP}, 06, 017,
  [\href{https://arxiv.org/abs/1002.2397}{1002.2397}].

\bibitem[{East {et~al.}(2018)East, Wojtak, \& Abel}]{East:2017qmk}
East, W.~E., Wojtak, R., \& Abel, T. 2018,
  \href{https://doi.org/10.1103/PhysRevD.97.043509}{Phys. Rev. D}, 97, 043509,
  [\href{https://arxiv.org/abs/1711.06681}{1711.06681}].

\bibitem[{Ellis(1984)}]{Ellis1984}
Ellis, G. F.~R. 1984, Relativistic Cosmology: Its Nature, Aims and Problems,
  ed. B.~Bertotti, F.~de~Felice, \& A.~Pascolini (Dordrecht: Springer
  Netherlands), 215--288

\bibitem[{Enqvist(2008)}]{Enqvist:2007vb}
Enqvist, K. 2008, \href{https://doi.org/10.1007/s10714-007-0553-9}{Gen. Rel.
  Grav.}, 40, 451, [\href{https://arxiv.org/abs/0709.2044}{0709.2044}].

\bibitem[{February {et~al.}(2014)February, Larena, Clarkson, \&
  Pollney}]{February:2013qza}
February, S., Larena, J., Clarkson, C., \& Pollney, D. 2014,
  \href{https://doi.org/10.1088/0264-9381/31/17/175008}{Class. Quant. Grav.},
  31, 175008, [\href{https://arxiv.org/abs/1311.5241}{1311.5241}].

\bibitem[{Fidler {et~al.}(2017)Fidler, Tram, Rampf, Crittenden, Koyama, \&
  Wands}]{Fidler:2017pnb}
Fidler, C., Tram, T., Rampf, C., {et~al.} 2017,
  \href{https://doi.org/10.1088/1475-7516/2017/12/022}{JCAP}, 12, 022,
  [\href{https://arxiv.org/abs/1708.07769}{1708.07769}].

\bibitem[{Flender {et~al.}(2013)Flender, Hotchkiss, \&
  Nadathur}]{Flender:2012wu}
Flender, S., Hotchkiss, S., \& Nadathur, S. 2013,
  \href{https://doi.org/10.1088/1475-7516/2013/02/013}{JCAP}, 02, 013,
  [\href{https://arxiv.org/abs/1212.0776}{1212.0776}].

\bibitem[{Garcia-Bellido \& Haugboelle(2008)}]{GarciaBellido:2008gd}
Garcia-Bellido, J. \& Haugboelle, T. 2008,
  \href{https://doi.org/10.1088/1475-7516/2008/09/016}{JCAP}, 09, 016,
  [\href{https://arxiv.org/abs/0807.1326}{0807.1326}].

\bibitem[{Garcia-Bellido \& Haugboelle(2009)}]{GarciaBellido:2008yq}
Garcia-Bellido, J. \& Haugboelle, T. 2009,
  \href{https://doi.org/10.1088/1475-7516/2009/09/028}{JCAP}, 0909, 028,
  [\href{https://arxiv.org/abs/0810.4939}{0810.4939}].

\bibitem[{Giblin {et~al.}(2016)Giblin, Mertens, \& Starkman}]{Giblin:2015vwq}
Giblin, J.~T., Mertens, J.~B., \& Starkman, G.~D. 2016,
  \href{https://doi.org/10.1103/PhysRevLett.116.251301}{Phys. Rev. Lett.}, 116,
  251301, [\href{https://arxiv.org/abs/1511.01105}{1511.01105}].

\bibitem[{Giblin {et~al.}(2019)Giblin, Mertens, Starkman, \&
  Tian}]{Giblin:2018ndw}
Giblin, J.~T., Mertens, J.~B., Starkman, G.~D., \& Tian, C. 2019,
  \href{https://doi.org/10.1103/PhysRevD.99.023527}{Phys. Rev. D}, 99, 023527,
  [\href{https://arxiv.org/abs/1810.05203}{1810.05203}].

\bibitem[{Green \& Wald(2014)}]{Green:2014aga}
Green, S.~R. \& Wald, R.~M. 2014,
  \href{https://doi.org/10.1088/0264-9381/31/23/234003}{Class. Quant. Grav.},
  31, 234003, [\href{https://arxiv.org/abs/1407.8084}{1407.8084}].

\bibitem[{Hui \& Greene(2006)}]{Hui:2005nm}
Hui, L. \& Greene, P.~B. 2006,
  \href{https://doi.org/10.1103/PhysRevD.73.123526}{Phys. Rev. D}, 73, 123526,
  [\href{https://arxiv.org/abs/astro-ph/0512159}{astro-ph/0512159}].

\bibitem[{Kaiser(2014)}]{kaiserone}
Kaiser, N. 2014, Elements of Astrophysics (CreateSpace Independent Publishing
  Platform)

\bibitem[{Kolb {et~al.}(2010)Kolb, Marra, \& Matarrese}]{Kolb:2009rp}
Kolb, E.~W., Marra, V., \& Matarrese, S. 2010,
  \href{https://doi.org/10.1007/s10714-009-0913-8}{Gen. Rel. Grav.}, 42, 1399,
  [\href{https://arxiv.org/abs/0901.4566}{0901.4566}].

\bibitem[{Laurent {et~al.}(2016)}]{Laurent:2016eqo}
Laurent, P. {et~al.} 2016,
  \href{https://doi.org/10.1088/1475-7516/2016/11/060}{JCAP}, 11, 060,
  [\href{https://arxiv.org/abs/1602.09010}{1602.09010}].

\bibitem[{Lavinto {et~al.}(2013)Lavinto, R\"as\"anen, \&
  Szybka}]{Lavinto:2013exa}
Lavinto, M., R\"as\"anen, S., \& Szybka, S.~J. 2013,
  \href{https://doi.org/10.1088/1475-7516/2013/12/051}{JCAP}, 12, 051,
  [\href{https://arxiv.org/abs/1308.6731}{1308.6731}].

\bibitem[{Leroy {et~al.}(2021)Leroy, Garrison, Eisenstein, Joyce, \&
  Maleubre}]{Leroy:2020fzc}
Leroy, M., Garrison, L., Eisenstein, D., Joyce, M., \& Maleubre, S. 2021,
  \href{https://doi.org/10.1093/mnras/staa3435}{Mon. Not. Roy. Astron. Soc.},
  501, 5064, [\href{https://arxiv.org/abs/2004.08406}{2004.08406}].

\bibitem[{Lim {et~al.}(2013)Lim, Regis, \& Clarkson}]{Lim:2013rra}
Lim, W.~C., Regis, M., \& Clarkson, C. 2013,
  \href{https://doi.org/10.1088/1475-7516/2013/10/010}{JCAP}, 10, 010,
  [\href{https://arxiv.org/abs/1308.0902}{1308.0902}].

\bibitem[{Macpherson {et~al.}(2019)Macpherson, Price, \&
  Lasky}]{Macpherson:2018btl}
Macpherson, H.~J., Price, D.~J., \& Lasky, P.~D. 2019,
  \href{https://doi.org/10.1103/PhysRevD.99.063522}{Phys. Rev. D}, 99, 063522,
  [\href{https://arxiv.org/abs/1807.01711}{1807.01711}].

\bibitem[{Marra {et~al.}(2007)Marra, Kolb, Matarrese, \& Riotto}]{Marra:2007pm}
Marra, V., Kolb, E.~W., Matarrese, S., \& Riotto, A. 2007,
  \href{https://doi.org/10.1103/PhysRevD.76.123004}{Phys. Rev. D}, 76, 123004,
  [\href{https://arxiv.org/abs/0708.3622}{0708.3622}].

\bibitem[{Marra \& Notari(2011)}]{Marra:2011ct}
Marra, V. \& Notari, A. 2011,
  \href{https://doi.org/10.1088/0264-9381/28/16/164004}{Class.Quant.Grav.}, 28,
  164004, [\href{https://arxiv.org/abs/1102.1015}{1102.1015}].

\bibitem[{Marra \& Paakkonen(2012)}]{Marra:2011zp}
Marra, V. \& Paakkonen, M. 2012,
  \href{https://doi.org/10.1088/1475-7516/2012/01/025}{JCAP}, 1201, 025,
  [\href{https://arxiv.org/abs/1105.6099}{1105.6099}].

\bibitem[{Matarrese {et~al.}(1998)Matarrese, Mollerach, \&
  Bruni}]{Matarrese:1997ay}
Matarrese, S., Mollerach, S., \& Bruni, M. 1998,
  \href{https://doi.org/10.1103/PhysRevD.58.043504}{Phys. Rev. D}, 58, 043504,
  [\href{https://arxiv.org/abs/astro-ph/9707278}{astro-ph/9707278}].

\bibitem[{{Matarrese} {et~al.}(1993){Matarrese}, {Pantano}, \&
  {Saez}}]{1993PhRvD..47.1311M}
{Matarrese}, S., {Pantano}, O., \& {Saez}, D. 1993,
  \href{https://doi.org/10.1103/PhysRevD.47.1311}{\prd}, 47, 1311.

\bibitem[{Meyer {et~al.}(2015)Meyer, Redlich, \& Bartelmann}]{Meyer:2014qla}
Meyer, S., Redlich, M., \& Bartelmann, M. 2015,
  \href{https://doi.org/10.1088/1475-7516/2015/03/053}{JCAP}, 03, 053,
  [\href{https://arxiv.org/abs/1412.3012}{1412.3012}].

\bibitem[{Michaux {et~al.}(2020)Michaux, Hahn, Rampf, \&
  Angulo}]{Michaux:2020yis}
Michaux, M., Hahn, O., Rampf, C., \& Angulo, R.~E. 2020,
  \href{https://doi.org/10.1093/mnras/staa3149}{Mon. Not. Roy. Astron. Soc.},
  500, 663, [\href{https://arxiv.org/abs/2008.09588}{2008.09588}].

\bibitem[{Moss {et~al.}(2011)Moss, Zibin, \& Scott}]{Moss:2010jx}
Moss, A., Zibin, J.~P., \& Scott, D. 2011,
  \href{https://doi.org/10.1103/PhysRevD.83.103515}{Phys. Rev. D}, 83, 103515,
  [\href{https://arxiv.org/abs/1007.3725}{1007.3725}].

\bibitem[{Nadathur {et~al.}(2012)Nadathur, Hotchkiss, \&
  Sarkar}]{Nadathur:2011iu}
Nadathur, S., Hotchkiss, S., \& Sarkar, S. 2012,
  \href{https://doi.org/10.1088/1475-7516/2012/06/042}{JCAP}, 06, 042,
  [\href{https://arxiv.org/abs/1109.4126}{1109.4126}].

\bibitem[{Nadathur {et~al.}(2014)Nadathur, Lavinto, Hotchkiss, \&
  R\"as\"anen}]{Nadathur:2014tfa}
Nadathur, S., Lavinto, M., Hotchkiss, S., \& R\"as\"anen, S. 2014,
  \href{https://doi.org/10.1103/PhysRevD.90.103510}{Phys. Rev. D}, 90, 103510,
  [\href{https://arxiv.org/abs/1408.4720}{1408.4720}].

\bibitem[{Nishikawa {et~al.}(2012)Nishikawa, Yoo, \& Nakao}]{Nishikawa:2012we}
Nishikawa, R., Yoo, C.-M., \& Nakao, K.-i. 2012,
  \href{https://doi.org/10.1103/PhysRevD.85.103511}{Phys. Rev. D}, 85, 103511,
  [\href{https://arxiv.org/abs/1202.1582}{1202.1582}].

\bibitem[{Ntelis {et~al.}(2017)}]{Ntelis:2017nrj}
Ntelis, P. {et~al.} 2017,
  \href{https://doi.org/10.1088/1475-7516/2017/06/019}{JCAP}, 06, 019,
  [\href{https://arxiv.org/abs/1702.02159}{1702.02159}].

\bibitem[{Perivolaropoulos \& Skara(2021)}]{Perivolaropoulos:2021jda}
Perivolaropoulos, L. \& Skara, F. ,
  [\href{https://arxiv.org/abs/2105.05208}{2105.05208}].

\bibitem[{Perlmutter {et~al.}(1999)}]{SupernovaCosmologyProject:1998vns}
Perlmutter, S. {et~al.} 1999, \href{https://doi.org/10.1086/307221}{Astrophys.
  J.}, 517, 565,
  [\href{https://arxiv.org/abs/astro-ph/9812133}{astro-ph/9812133}].

\bibitem[{{Ragagnin} {et~al.}(2020){Ragagnin}, {Dolag}, {Wagner}, {Gheller},
  {Roffler}, {Goz}, {Hubber}, \& {Arth}}]{2020arXiv200310850R}
{Ragagnin}, A., {Dolag}, K., {Wagner}, M., {et~al.} 2020, in Advances in
  Parallel Computing, Vol.~36, Parallel Computing: Technology Trends, ed.
  I.~{Foster} {et~al.}, 209--218

\bibitem[{Rasanen(2009)}]{Rasanen:2009mg}
Rasanen, S. 2009, \href{https://doi.org/10.1103/PhysRevD.79.123522}{Phys. Rev.
  D}, 79, 123522, [\href{https://arxiv.org/abs/0903.3013}{0903.3013}].

\bibitem[{Redlich {et~al.}(2014)Redlich, Bolejko, Meyer, Lewis, \&
  Bartelmann}]{Redlich:2014gga}
Redlich, M., Bolejko, K., Meyer, S., Lewis, G.~F., \& Bartelmann, M. 2014,
  \href{https://doi.org/10.1051/0004-6361/201424553}{Astron. Astrophys.}, 570,
  A63, [\href{https://arxiv.org/abs/1408.1872}{1408.1872}].

\bibitem[{Riess {et~al.}(1998)}]{SupernovaSearchTeam:1998fmf}
Riess, A.~G. {et~al.} 1998, \href{https://doi.org/10.1086/300499}{Astron. J.},
  116, 1009, [\href{https://arxiv.org/abs/astro-ph/9805201}{astro-ph/9805201}].

\bibitem[{Riess {et~al.}(2021)}]{Riess:2021jrx}
Riess, A.~G. {et~al.} , [\href{https://arxiv.org/abs/2112.04510}{2112.04510}].

\bibitem[{Rigopoulos \& Valkenburg(2012)}]{Rigopoulos:2012xj}
Rigopoulos, G. \& Valkenburg, W. 2012,
  \href{https://doi.org/10.1103/PhysRevD.86.043523}{Phys. Rev. D}, 86, 043523,
  [\href{https://arxiv.org/abs/1203.2796}{1203.2796}].

\bibitem[{{Sachs} \& {Wolfe}(1967)}]{1967ApJ...147...73S}
{Sachs}, R.~K. \& {Wolfe}, A.~M. 1967,
  \href{https://doi.org/10.1086/148982}{\apj}, 147, 73.

\bibitem[{Sakai \& Inoue(2008)}]{Sakai:2008fi}
Sakai, N. \& Inoue, K.~T. 2008,
  \href{https://doi.org/10.1103/PhysRevD.78.063510}{Phys. Rev. D}, 78, 063510,
  [\href{https://arxiv.org/abs/0805.3446}{0805.3446}].

\bibitem[{Scrimgeour {et~al.}(2012)}]{Scrimgeour:2012wt}
Scrimgeour, M. {et~al.} 2012,
  \href{https://doi.org/10.1111/j.1365-2966.2012.21402.x}{Mon. Not. Roy.
  Astron. Soc.}, 425, 116, [\href{https://arxiv.org/abs/1205.6812}{1205.6812}].

\bibitem[{{Silk}(1977)}]{1977A&A....59...53S}
{Silk}, J. 1977, \aap, 59, 53.

\bibitem[{Springel(2005)}]{Springel:2005mi}
Springel, V. 2005, \href{https://doi.org/10.1111/j.1365-2966.2005.09655.x}{Mon.
  Not. Roy. Astron. Soc.}, 364, 1105,
  [\href{https://arxiv.org/abs/astro-ph/0505010}{astro-ph/0505010}].

\bibitem[{Springel {et~al.}(2021)Springel, Pakmor, Zier, \&
  Reinecke}]{Springel:2020plp}
Springel, V., Pakmor, R., Zier, O., \& Reinecke, M. 2021,
  \href{https://doi.org/10.1093/mnras/stab1855}{Mon. Not. Roy. Astron. Soc.},
  506, 2871, [\href{https://arxiv.org/abs/2010.03567}{2010.03567}].

\bibitem[{Stebbins(2012)}]{Stebbins:2012vw}
Stebbins, A. 2012, \href{https://doi.org/10.1142/S0218271812420175}{Int. J.
  Mod. Phys. D}, 21, 1242017,
  [\href{https://arxiv.org/abs/1205.4201}{1205.4201}].

\bibitem[{Sussman(2011)}]{Sussman:2011na}
Sussman, R.~A. 2011,
  \href{https://doi.org/10.1088/0264-9381/28/23/235002}{Class. Quant. Grav.},
  28, 235002, [\href{https://arxiv.org/abs/1102.2663}{1102.2663}].

\bibitem[{{Szekeres}(1975)}]{1975CMaPh..41...55S}
{Szekeres}, P. 1975, \href{https://doi.org/10.1007/BF01608547}{Communications
  in Mathematical Physics}, 41, 55.

\bibitem[{{Taffoni} {et~al.}(2020){Taffoni}, {Becciani}, {Garilli}, {Maggio},
  {Pasian}, {Umana}, {Smareglia}, \& {Vitello}}]{2020ASPC..527..307T}
{Taffoni}, G., {Becciani}, U., {Garilli}, B., {et~al.} 2020, in Astronomical
  Society of the Pacific Conference Series, ed. R.~{Pizzo}, E.~R. {Deul}, J.~D.
  {Mol}, J.~{de Plaa}, \& H.~{Verkouter}, Vol. 527, 307

\bibitem[{Valkenburg(2012{\natexlab{a}})}]{Valkenburg:2011tm}
Valkenburg, W. 2012{\natexlab{a}},
  \href{https://doi.org/10.1007/s10714-012-1405-9}{Gen.Rel.Grav.}, 44, 2449,
  [\href{https://arxiv.org/abs/1104.1082}{1104.1082}].

\bibitem[{Valkenburg(2012{\natexlab{b}})}]{Valkenburg:2011ty}
Valkenburg, W. 2012{\natexlab{b}},
  \href{https://doi.org/10.1088/1475-7516/2012/01/047}{JCAP}, 01, 047,
  [\href{https://arxiv.org/abs/1106.6042}{1106.6042}].

\bibitem[{Valkenburg \& Hu(2015)}]{Valkenburg:2015dsa}
Valkenburg, W. \& Hu, B. 2015,
  \href{https://doi.org/10.1088/1475-7516/2015/09/054}{JCAP}, 09, 054,
  [\href{https://arxiv.org/abs/1505.05865}{1505.05865}].

\bibitem[{Valkenburg {et~al.}(2014)Valkenburg, Marra, \&
  Clarkson}]{Valkenburg:2012td}
Valkenburg, W., Marra, V., \& Clarkson, C. 2014,
  \href{https://doi.org/10.1093/mnrasl/slt140}{Mon. Not. Roy. Astron. Soc.},
  438, L6, [\href{https://arxiv.org/abs/1209.4078}{1209.4078}].

\bibitem[{Valkenburg \& Villaescusa-Navarro(2017)}]{Valkenburg:2016xek}
Valkenburg, W. \& Villaescusa-Navarro, F. 2017,
  \href{https://doi.org/10.1093/mnras/stx376}{Mon. Not. Roy. Astron. Soc.},
  467, 4401, [\href{https://arxiv.org/abs/1610.08501}{1610.08501}].

\bibitem[{Van~Acoleyen(2008)}]{VanAcoleyen:2008cy}
Van~Acoleyen, K. 2008,
  \href{https://doi.org/10.1088/1475-7516/2008/10/028}{JCAP}, 10, 028,
  [\href{https://arxiv.org/abs/0808.3554}{0808.3554}].

\bibitem[{Vielva(2010)}]{Vielva:2010ng}
Vielva, P. 2010, \href{https://doi.org/10.1155/2010/592094}{Adv. Astron.},
  2010, 592094, [\href{https://arxiv.org/abs/1008.3051}{1008.3051}].

\bibitem[{{Villaescusa-Navarro}(2018)}]{2018ascl.soft11008V}
{Villaescusa-Navarro}, F. 2018, Astrophysics Source Code Library, 1811.008

\bibitem[{Wang \& Steinhardt(1998)}]{Wang:1998gt}
Wang, L.-M. \& Steinhardt, P.~J. 1998,
  \href{https://doi.org/10.1086/306436}{Astrophys. J.}, 508, 483,
  [\href{https://arxiv.org/abs/astro-ph/9804015}{astro-ph/9804015}].

\bibitem[{Yamamoto {et~al.}(2016)Yamamoto, Marra, Mukhanov, \&
  Sasaki}]{Yamamoto:2015etj}
Yamamoto, K., Marra, V., Mukhanov, V., \& Sasaki, M. 2016,
  \href{https://doi.org/10.1088/1475-7516/2016/03/030}{JCAP}, 03, 030,
  [\href{https://arxiv.org/abs/1512.04240}{1512.04240}].

\bibitem[{Zhang \& Stebbins(2011)}]{Zhang:2010fa}
Zhang, P. \& Stebbins, A. 2011,
  \href{https://doi.org/10.1103/PhysRevLett.107.041301}{Phys.Rev.Lett.}, 107,
  041301, [\href{https://arxiv.org/abs/1009.3967}{1009.3967}].

\bibitem[{Zibin(2008)}]{Zibin:2008vj}
Zibin, J.~P. 2008, \href{https://doi.org/10.1103/PhysRevD.78.043504}{Phys.
  Rev.}, D78, 043504, [\href{https://arxiv.org/abs/0804.1787}{0804.1787}].

\bibitem[{Zibin(2011)}]{Zibin:2011ma}
Zibin, J.~P. 2011,
  \href{https://doi.org/10.1103/PhysRevD.84.123508}{Phys.Rev.}, D84, 123508,
  [\href{https://arxiv.org/abs/1108.3068}{1108.3068}].

\bibitem[{Zibin(2014)}]{Zibin:2014vaa}
Zibin, J.~P. , [\href{https://arxiv.org/abs/1408.4442}{1408.4442}].

\bibitem[{Zibin \& Moss(2011)}]{Moss:2011ze}
Zibin, J.~P. \& Moss, A. 2011,
  \href{https://doi.org/10.1088/0264-9381/28/16/164005}{Class.Quant.Grav.}, 28,
  164005, [\href{https://arxiv.org/abs/1105.0909}{1105.0909}].

\end{thebibliography}

\begin{appendix}

\section{Latest constraints on $\Lambda$LTB}
\label{ap:constraints}

\begin{figure}[t!]
\centering 
\includegraphics[trim={.cm .cm .cm .cm}, clip, width= \columnwidth]{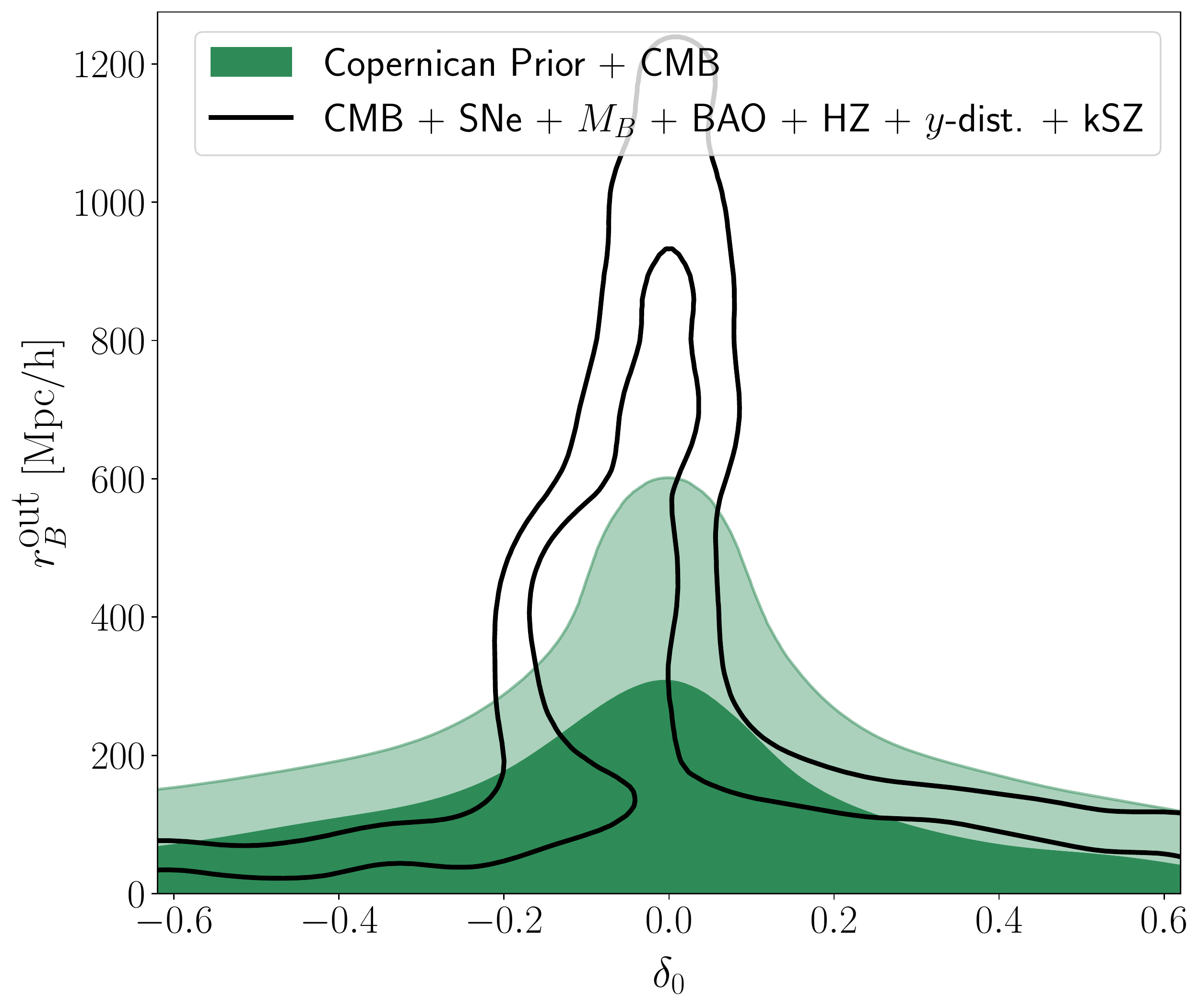}
\caption{
 {Marginalized constraints on the central contrast $\delta_0$ and inhomogeneous size $r_b$ of the $\Lambda$LTB model at 68\% and 95\% confidence level. The empty contours show the constraints from CMB, BAO, type Ia supernovae, local $H_0$, cosmic chronometers, Compton y-distortion and kinetic Sunyaev–Zeldovich effect.
The green area shows the region of the parameter space that is allowed by the standard model, here represented via the Copernican prior convolved with the CMB likelihood. See \citet[]{Camarena:2021mjr} for more details.}
\label{fig:constraints}}
\end{figure}

 {Here, we show the latest constraints on the $\Lambda$LTB model for the case of the observer at the center of the spherical inhomogeneity, as obtained in \citet[]{Camarena:2021mjr}.
In Figure~\ref{fig:constraints}, we report the constraints using the same LTB parameters that are adopted here, that is, the central contrast $\delta_0$ and inhomogeneous size $r_b$.
We see that contrasts $|\delta_0| \gtrsim 0.2$ are ruled out at scales $r_b \gtrsim 300$ Mpc/$h$. Note, however, that \citet[]{Camarena:2021mjr} adopted a different curvature profile with respect to the one of Eq.~\eqref{profi1}.
}

\section{O\lowercase{pen}G\lowercase{adget3} performance}
\label{ap:perfo}

\begin{figure*}
\includegraphics[width= \textwidth]{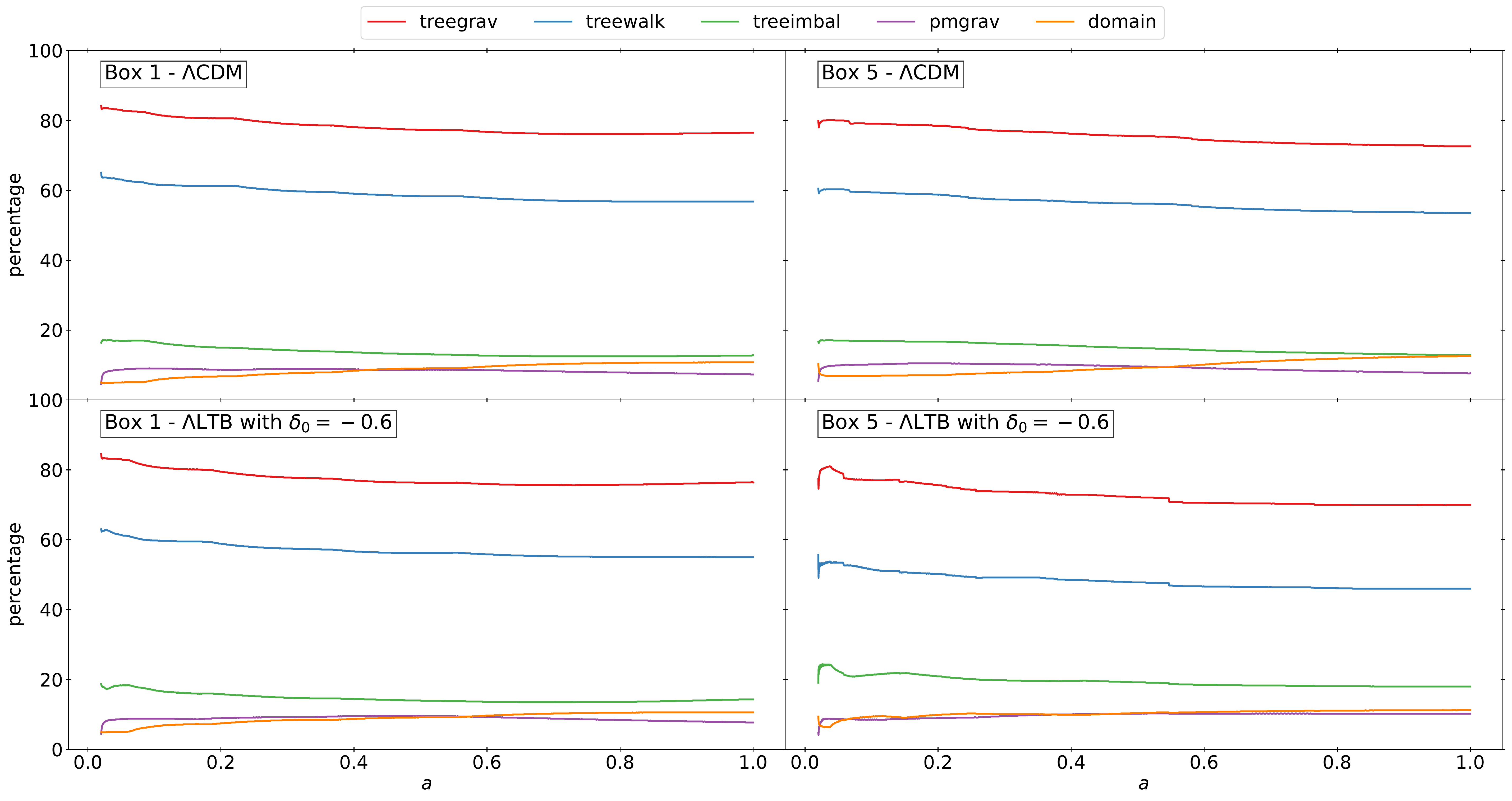}
\caption{
Logs of two $\Lambda$LTB simulations (bottom) and the corresponding $\Lambda$CDM ones (top) executed with \texttt{OpenGadget3}.\label{fig:perfo}}
\end{figure*}

Figure~\ref{fig:perfo} illustrates how the execution of a simulation with \texttt{OpenGadget3} is affected by the background inhomogeneity.
We show the CPU times of the various internal algorithms, as a function of the scale factor, for the most nonlinear $\Lambda$LTB simulations ($\delta_0=-0.6$, bottom) and the corresponding $\Lambda$CDM ones (top).
One can see a very similar behavior, with a small increase in tree imbalance  for the $\Lambda$LTB simulation.

For all simulations, the parameters that control the structure of the gravity solver are set to $A_{\rm smth}=1.25$ and $R_{\rm cut}=4.5$.
$A_{\rm smth}$ sets the scale in units of mesh-cells that defines the long-range/short-range force-split in the TreePM algorithm. A larger value of $A_{\rm smth}$ will make the transition region better resolved by the mesh, yielding higher accuracy and less residual scatter in the force matching region, but at the same time the region that needs to be covered by the tree grows, which makes the computation more expensive.
$R_{\rm cut}$ sets the maximum radius out to which the short-range tree-force is evaluated in case the TreePM algorithm is used.

We also tested the performance of the GPU porting of \texttt{OpenGadget3} using OpenACC directives that was presented in \citet[]{2020arXiv200310850R}. We found that the GPU porting is $\approx$40\% faster when using $256^3$ particles and $256^3$ PM elements, but that the performance is similar when using $256^3$ particles and $512^3$ PM elements and that it is $\approx$20\% slower when using  $512^3$ particles and $1024^3$ PM elements. From the log files one can see that the `pmgrav' module takes  $\approx$30\% resources up to $z=1.5$ and $\approx$20\% up to $z=0$.
Given these preliminary results, we used the version of \texttt{OpenGadget3} without GPUs.

\section{Validation of the $\Lambda$CDM simulations}
\label{ap:LCDMtest}


\begin{figure}
\centering 
\includegraphics[trim={.cm .cm .cm .cm}, clip, width= \columnwidth]{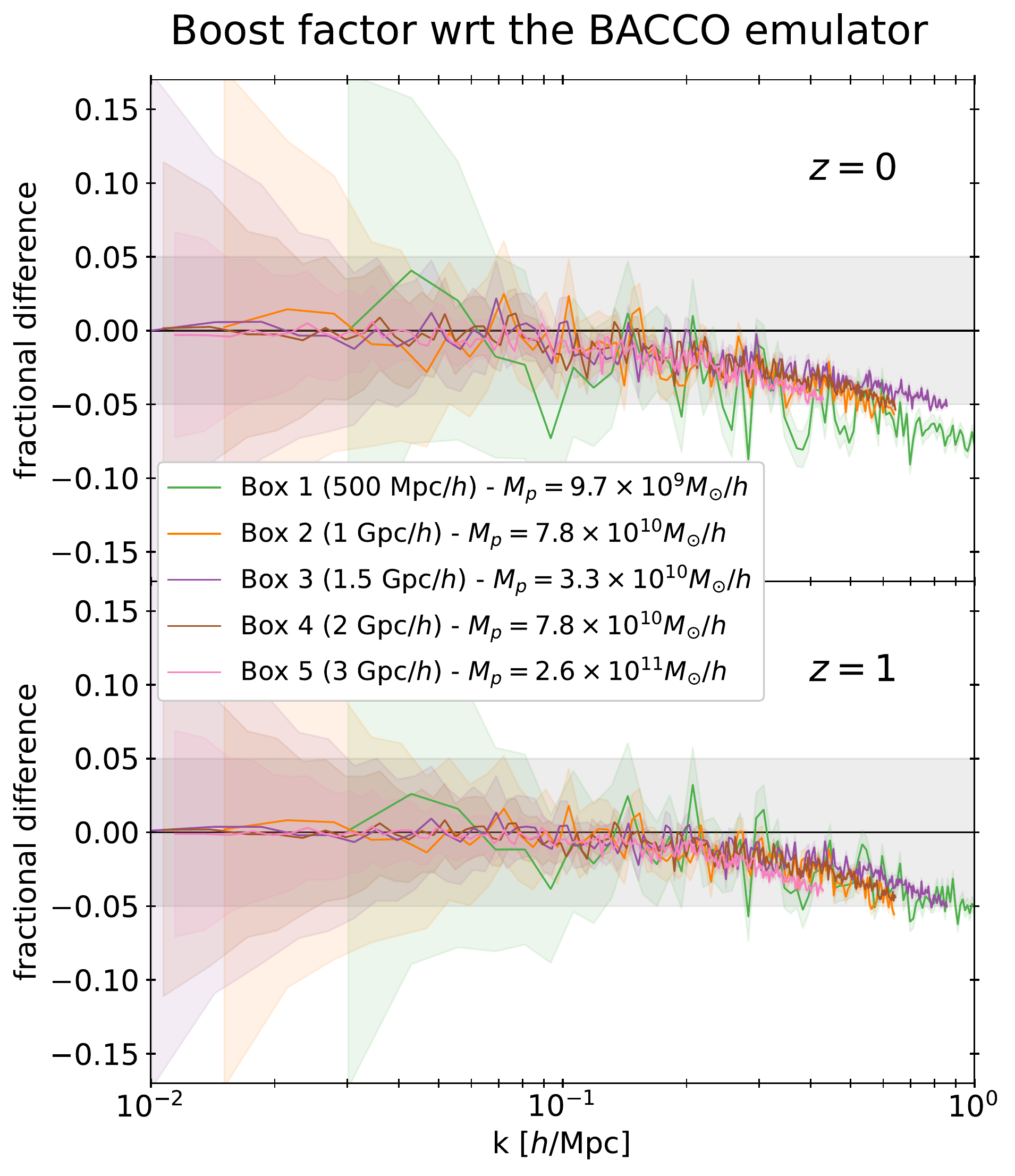}
\caption{
Boost factor $P(k,z)/P(k,z=49)$ as compared with the one from the BACCO emulator (top for $z=0$, bottom for $z=1$).
\label{fig:LCDMtest-pk}}
\end{figure}

\begin{figure}
\centering 
\includegraphics[trim={.cm .cm .cm .cm}, clip, width= \columnwidth]{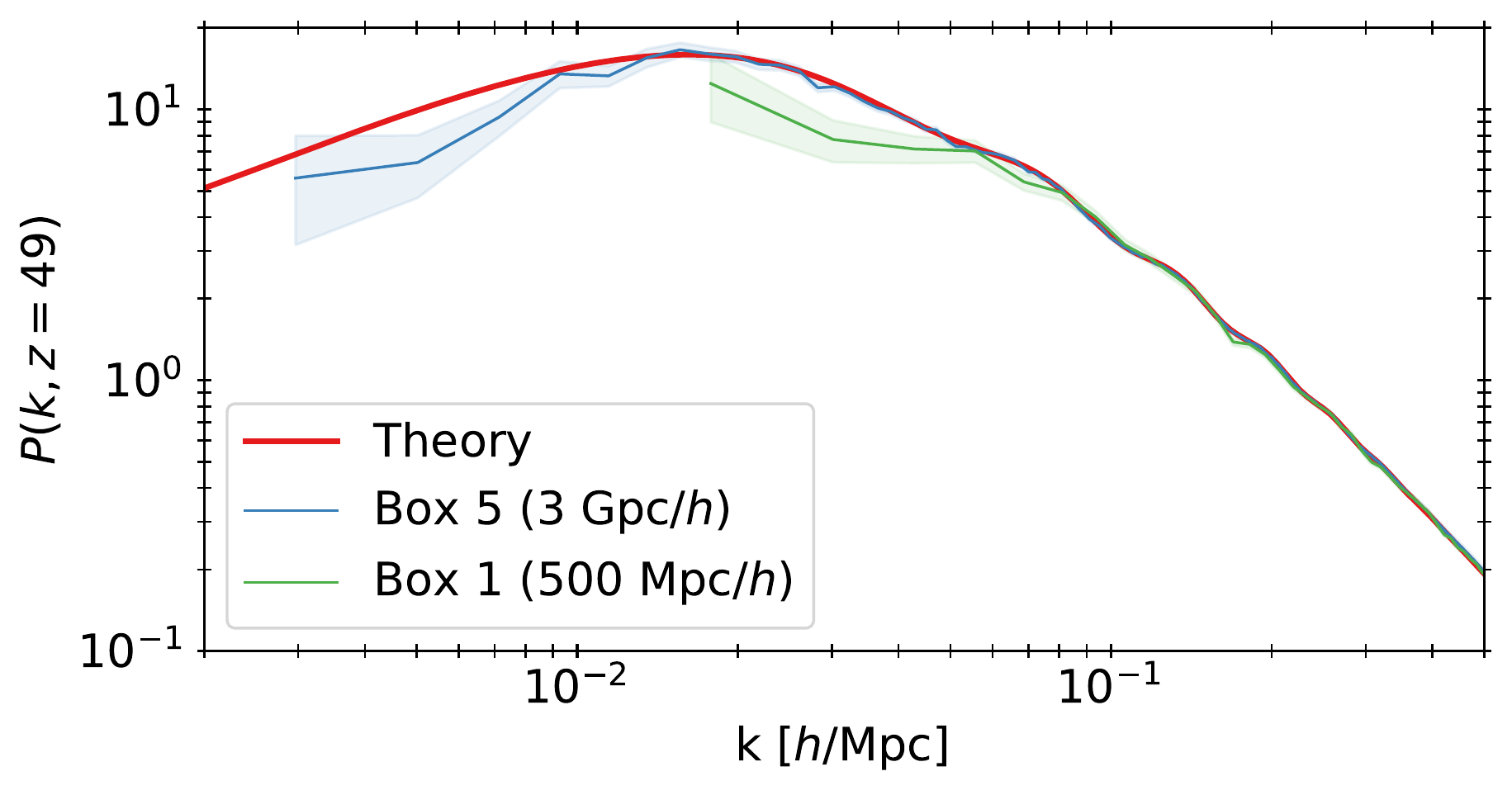}
\caption{
Initial measured power spectrum for Box 1 and Box 5.
The first few modes have less power with respect to the theoretical spectrum (we adopted the same seed for all  simulations).
\label{fig:LCDMtest-pk49}}
\end{figure}

Here, we validate the $\Lambda$CDM simulations against the BACCO emulator \citep[v2.1.0,][]{Angulo:2020vky} and the halo mass function of \citet{Castrone}. 
The emulator is expected to be accurate at the 2\% level, and the Castro22 mass function strives to obtain a better than 1\% calibration for next-generation surveys.

First, we consider the power spectrum.
Figure~\ref{fig:LCDMtest-pk} compares the boost factor $P(k,z)/P(k,z=49)$ with the one from the BACCO emulator (top for $z=0$, bottom for $z=1$).
We estimate the power spectrum with \texttt{Pylians3}%
\footnote{\href{https://github.com/franciscovillaescusa/Pylians3}{github.com/franciscovillaescusa/Pylians3}}
\citep[][]{2018ascl.soft11008V} and consider wavenumbers till a tenth of the Nyquist wavenumber so that aliasing errors are  negligible.
Error bars are estimated via $\sigma_P=( P(k,z)+ 1/ n_v)/\sqrt{N_k}$ where $N_k$ is the number of independent modes used to estimate $P(k)$, and $ n_v$  is the number density in the simulation box.
We see that all boxes but the smallest Box 1 have better than 5\% accurate power spectrum.
We adopted the same seed for all the simulations and it happened that the first few modes have less power with respect to the theoretical spectrum, see Figure~\ref{fig:LCDMtest-pk49}.
In the case of Box 1, the first few modes are at a larger wavenumber so that the worse accuracy that we see in this case could be due to mode coupling.
Indeed, having less power on large scales in the initial conditions also implies that less power is transferred to smaller scales, due to the coupling between different modes during the nonlinear evolution.
Note, indeed, that the discrepancy is less severe at $z=1$ (Fig.~\ref{fig:LCDMtest-pk}, bottom panel).
We also performed a Box-2 simulation (1 Gpc and $N_{\rm part}=1024^3$, see Table~\ref{tab:sims}) with the cosmology of \citet{Castrone} and compared it with a 1-Gpc simulation with $N_{\rm part}=4064^3$ that was run according to the specifications of \citet{Castrone}.
We found an agreement better than 5\%.

\begin{figure}
\centering 
\includegraphics[width= \columnwidth]{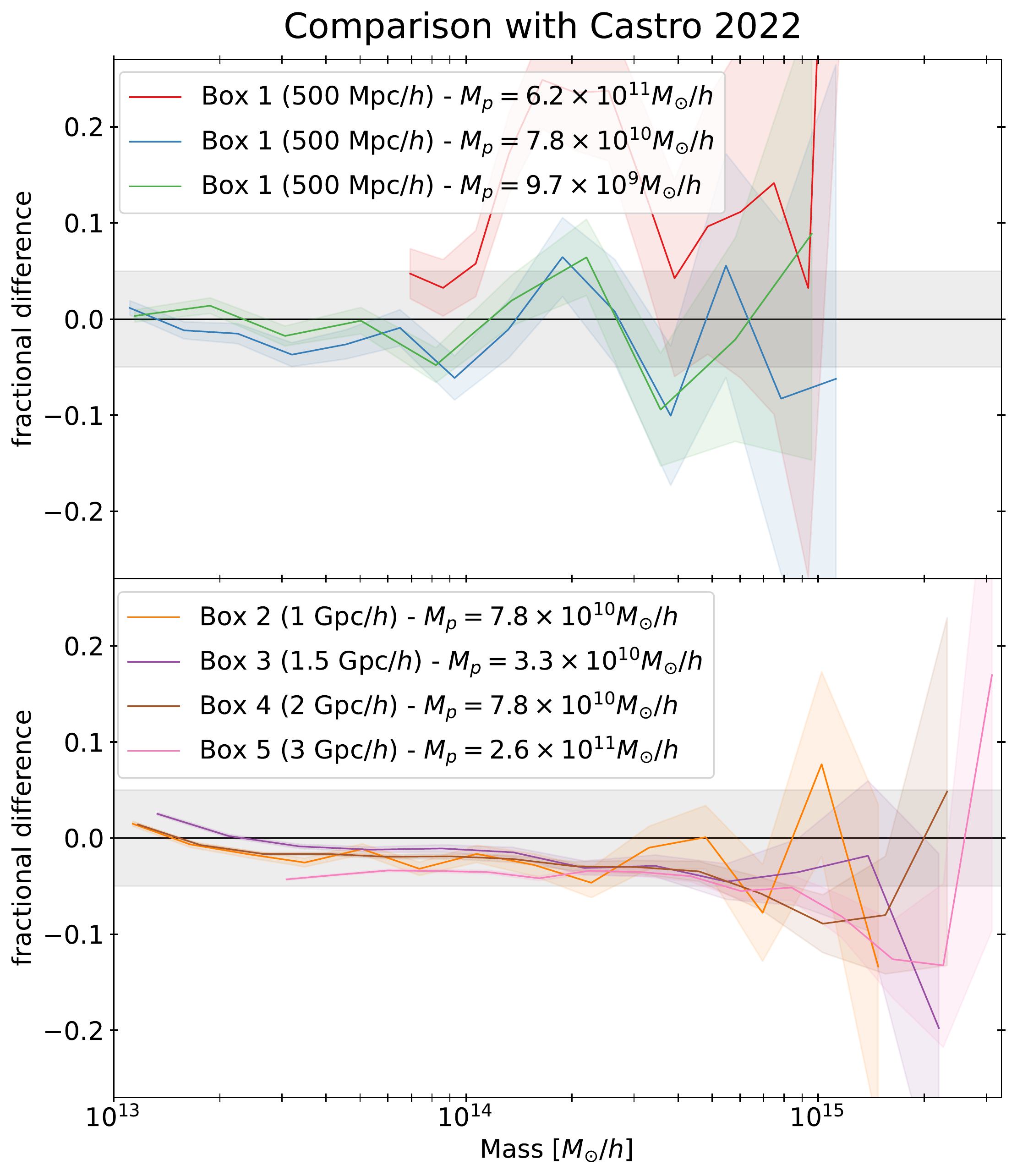}
\caption{
Halo mass function from the simulations against the one of 
\citet{Castrone}, which adopts the virial spherical overdensity.
See Appendix~\ref{ap:LCDMtest} for more details.
\label{fig:LCDMtest-hmf}}
\end{figure}

Next, we consider halo abundances.
Figure~\ref{fig:LCDMtest-hmf} compares the halo mass function against the one of 
\citet{Castrone}, 
which adopts the virial spherical overdensity.
We adopt  Gaussian error bars from the corresponding Poisson distributions.
The top panel shows the fractional difference for Box 1 and increasing mass resolution, that is, $256^3$, $512^3$ and $1024^3$ particles. One sees that with a particle mass of $\lesssim 10^{11} M_\odot /h$ one reaches a 5\% accuracy. Then we show, in the bottom panel, the fractional difference for the other boxes, again reaching a 5\% accuracy.

Concluding, the $\Lambda$CDM simulations of this first set of BEHOMO simulations have power spectrum and halo mass function accurate at the 5\% level (except Box 1).
Note that the $\Lambda$LTB and $\Lambda$CDM simulations are paired so that numerical errors should approximately factor out when considering suitable ratios of relevant quantities.

\end{appendix}

\end{document}